\documentclass[twocolumn,superscriptaddress,secnumarabic,
amssymb,amsmath,nobibnotes,aps,prd,showpacs,nofootinbib]{revtex4}%
\usepackage{graphicx}
\usepackage{epsf}
\usepackage{bm}
\usepackage{amsmath}
\usepackage{amsfonts}
\usepackage{amssymb}
\usepackage{epstopdf}
\usepackage{natbib}%
\usepackage{rotating}
\usepackage{pdflscape}
\usepackage{adjustbox}
\providecommand{\U}[1]{\protect\rule{.1in}{.1in}}
\newcommand{\be}{\begin{equation}}
\newcommand{\ee}{\end{equation}}

\newcommand{\mincir}{\raise
-3.truept\hbox{\rlap{\hbox{$\sim$}}\raise4.truept\hbox{$<$}\ }}
\newcommand{\magcir}{\raise
-3.truept\hbox{\rlap{\hbox{$\sim$}}\raise4.truept\hbox{$>$}\ }}

\usepackage{color}

\begin{document}
\title{Dynamical dark energy after Planck CMB final release and $H_0$ tension}

\author{Weiqiang Yang}
\email{d11102004@163.com}
\affiliation{Department of Physics, Liaoning Normal University, Dalian, 116029, P. R. China}

\author{Eleonora Di Valentino}
\email{eleonora.divalentino@manchester.ac.uk}
\affiliation{Jodrell Bank Center for Astrophysics, School of Physics and Astronomy, 
University of Manchester, Oxford Road, Manchester, M13 9PL, UK.}

\author{Supriya Pan}
\email{supriya.maths@presiuniv.ac.in}
\affiliation{Department of Mathematics, Presidency University, 86/1 College Street, 
Kolkata 700073, India.}

\author{Yabo Wu}
\email{ybwu61@163.com}
\affiliation{Department of Physics, Liaoning Normal University, Dalian, 116029, P. R. China}

\author{Jianbo Lu}
\email{lvjianbo819@163.com}
\affiliation{Department of Physics, Liaoning Normal University, Dalian, 116029, P. R. China}



\begin{abstract}
In this article we compare a variety of well known dynamical dark energy models using the cosmic microwave background measurements from the 2018 Planck legacy and 2015 Planck data releases, the baryon acoustic oscillations measurements and the local measurements of $H_0$ obtained by the SH0ES (Supernovae, $H_0$, for the Equation of State of Dark energy) collaboration analysing the Hubble Space Telescope data. We discuss the alleviation of $H_0$ tension, that is obtained at the price of a phantom-like dark energy equation of state. We perform a Bayesian evidence analysis to quantify the improvement of the fit, finding that all the dark energy models considered in this work are preferred against the $\Lambda$CDM scenario. Finally, among all the possibilities analyzed, the CPL model is the best one in fitting the data and solving the $H_0$ tension at the same time. However, unfortunately, this dynamical dark energy solution is not supported by the baryon acoustic oscillations (BAO) data, and the tension is restored when BAO data are included for all the models.  
\end{abstract}

\maketitle
\section{Introduction}

If we look at the discoveries within the boundary of cosmological physics, certainly, the late-time cosmic acceleration is one of them. It was first probed through the investigations of supernovae Type Ia  {\color{blue}\cite{Riess:1998cb,Perlmutter:1998np}}, and has then been confirmed by several complementary observations {\color{blue}\cite{Suzuki:2011hu,Crocce:2015xpb,Alam:2015mbd}}. Within the context of general theory of relativity, an exotic fluid having large negative pressure, dubbed as dark energy, is necessary for explaining this accelerating universe. Since the process is phenomenological, in the sense that we need some extra fluids with negative pressure, thus, it is very easy to propose a number of dark energy models, as has already been witnessed {\color{blue}\cite{Copeland:2006wr,Huterer:2017buf,Sergijenko:2011vb,Paliathanasis:2014yfa,Paliathanasis:2014zxa,Paliathanasis:2015gga,Gerardi:2019obr,Benaoum:2020qsi}}. 
According to the observational results, the $\Lambda$-Cold Dark Matter ($\Lambda$CDM) cosmology is probably the best fitted cosmological model in which the cosmological constant $\Lambda$ accelerates the expansion of our universe. The equation of state parameter of $\Lambda$, namely, $w_{\Lambda} =  p_{\Lambda}/\rho_{\lambda} = -1$, being negative, behaves like a dark energy fluid.

Despite its notable success in explaining the late-accelerating phase of the universe, at the same time, $\Lambda$CDM is equally suffering with some serious issues. Apart from the cosmological constant problem, i.e. the disagreement between the observed small value of the cosmological constant and the theoretical large value of vacuum suggested by the quantum field theory, the $\Lambda$-cosmology is also unable to address the cosmic coincidence problem. This motivated the scientific community to find some alternative cosmological models that might be free from the above two problems. On the other hand, within the regime of $\Lambda$-cosmology,  some of the key cosmological parameters, namely the Hubble constant $H_0$ and the amplitude of the matter power spectrum $\sigma_8$, differ unexpectedly with their
estimations by different astronomical missions at several standard deviations.

As an example, if we look at the estimation of the Hubble constant $H_0$ from $\Lambda$CDM based Planck's mission {\color{blue}\cite{Ade:2015xua}}, we see that this estimation is, (i) more than $4\sigma$ apart compared to its estimation by SH0ES collaboration~{\color{blue}\cite{Riess:2019cxk}}, (ii) more than $5\sigma$ if SH0ES is combined with the H0liCOW collaboration~{\color{blue}\cite{Wong:2019kwg}}, and (iii) around $4.7\sigma$ apart from the cosmology independent local determination~{\color{blue}\cite{Camarena:2019moy}}. On the other hand, the parameter $S_8~(\equiv \sigma_8 \sqrt{\Omega_{m0}/0.3})$ when measured from Planck~{\color{blue}\cite{Ade:2015xua}} within the assumption of a $\Lambda$CDM scenario, is many sigmas apart compared to its estimation by cosmic shear missions, namely,  KiDS-450~{\color{blue}\cite{Kuijken:2015vca,Hildebrandt:2016iqg,Conti:2016gav}}, DES~{\color{blue}\cite{Abbott:2017wau,Troxel:2017xyo}} and CFHTLenS~{\color{blue}\cite{Heymans:2012gg, Erben:2012zw,Joudaki:2016mvz}}.  Due to such discordances it was naturally thought that some new physical theory beyond the standard  $\Lambda$CDM  paradigm  {\color{blue}\cite{Mortsell:2018mfj,Vagnozzi:2019ezj}} could be essential. To discuss the possibility of solving or alleviating the $H_0$ tension (also see {\color{blue}\cite{Bernal:2016gxb,Garcia-Quintero:2019cgt,Wu:2020nxz}}), some of the proposed cosmological theories in this series are, dynamical dark energy {\color{blue}\cite{DiValentino:2016hlg,DiValentino:2017zyq,Yang:2018qmz,Li:2019yem,Pan:2019hac,DiValentino:2020naf,Banihashemi:2020wtb,Alestas:2020mvb}}, dark matter -- neutrino interactions {\color{blue}\cite{DiValentino:2017oaw}}, bulk viscosity {\color{blue}\cite{Yang:2019qza}}, unified cosmology {\color{blue}\cite{Yang:2019nhz,Yang:2019jwn}}, metastable dark energy {\color{blue}\cite{Li:2019san,Yang:2020zuk}}, vacuum phase transition {\color{blue}\cite{DiValentino:2017rcr,DiValentino:2020kha}}, modified gravity theory {\color{blue}\cite{Khosravi:2017hfi,Renk:2017rzu,Nunes:2018xbm,Cai:2019bdh,Wang:2020zfv}}, early dark energy {\color{blue}\cite{Karwal:2016vyq,Poulin:2018cxd,Niedermann:2020dwg,Sakstein:2019fmf}}, quintessence dark energy {\color{blue}\cite{DiValentino:2019exe}}, light dark matter {\color{blue}\cite{Alcaniz:2019kah}}, self interacting dark matter {\color{blue}\cite{Hryczuk:2020jhi}}, swampland {\color{blue}\cite{Colgain:2018wgk,Colgain:2019joh}}, Ginzburg-Landau Theory of Dark Energy {\color{blue}\cite{Banihashemi:2018has}}, hot axions {\color{blue}\cite{DEramo:2018vss}}, evolving scalar field {\color{blue}\cite{Agrawal:2019lmo}},  interacting radiation {\color{blue}\cite{Blinov:2020hmc}}, decaying dark matter {\color{blue}\cite{Vattis:2019efj,Clark:2020miy}}, ultra light scalar decay {\color{blue}\cite{Gonzalez:2020fdy}}, evolving gravitational constant {\color{blue}\cite{Braglia:2020iik}}, primordial magnetic fields {\color{blue}\cite{Jedamzik:2020krr}}, interacting dark energy {\color{blue}\cite{DiValentino:2017iww,Kumar:2016zpg,Kumar:2017dnp,Buen-Abad:2017gxg,Yang:2018euj,Yang:2018uae,Kumar:2019wfs,Martinelli:2019dau,Pan:2019gop,Yang:2019uog,Yang:2019uzo,Pan:2020bur}},  sterile neutrino self-interactions {\color{blue}\cite{Archidiacono:2020yey}}, Non-standard Recombination {\color{blue}\cite{Liu:2019awo}}, thermal friction {\color{blue}\cite{Berghaus:2019cls}}, screened fifth forces {\color{blue}\cite{Desmond:2019ygn}} etc. On the other hand, for the $S_8$ tension, one could see {\color{blue}\cite{Pourtsidou:2016ico,An:2017crg,DiValentino:2018gcu,Kazantzidis:2018rnb,Kumar:2019wfs}}.

The simplest generalization of the $\Lambda$CDM cosmology is the $w$CDM cosmology in which the equation of state parameter $w$ of the dark energy fluid is constant and $w \neq -1$. An even better generalization of the $\Lambda$CDM cosmology is to consider a time dependent equation of state parameter of the underlying dark energy fluid. This eventually gives a freedom to reconstruct the expansion history of the universe with the observational data. Following this approach, a cluster of $w(t)$ models have been proposed and tested with a variety of cosmological observations from different sources, see {\color{blue}\cite{Chevallier:2000qy,Linder:2002et,Jassal:2005qc,Efstathiou:1999tm,Barboza:2008rh,Zhao:2017cud,Yang:2017amu,DiValentino:2017gzb,Rezaei:2017yyj,Marcondes:2017vjw,Yang:2017alx,Vagnozzi:2018jhn}} and the references therein. This particular area of cosmological models got crucial attention after a recent work {\color{blue}\cite{Zhao:2017cud}} which argued that the observational data prefer the dynamical state parameter in the dark energy equation of state. Now, since observational cosmology is getting richer with the newly available potential cosmological datasets, therefore, it is essential to test the evolution of the dark energy equation of state for a convincing result. On the other hand, the viability of any specific dark energy equation of state can also be tested in light of the recently available cosmological probes. Therefore, in the present article we have considered some very well known dark energy models parametrized by their dynamical equation of state parameters and constrained their parameter space using the cosmic microwave background data from both Planck 2015 {\color{blue}\cite{Ade:2015xua,Adam:2015rua,Aghanim:2015xee}} and Planck 2018 data release {\color{blue}\cite{Aghanim:2018eyx,Aghanim:2018oex,Aghanim:2019ame}}. Apart from the cosmological constraints extracted out of the proposed dark energy parametrizations in this work, we aim to examine whether the parametrized dark energy models can resolve the Hubble constant tension problem.

The paper has been organized in the following way. In section \ref{sec-2} we describe the gravitational equations driven by the components of the universe 
and the models. Then in section \ref{sec-data} we discuss the observational data sets that we use to constrain the proposed dark energy models. After that in section \ref{sec-results} we describe the observational constraints on the parameter space of all the models. Finally, in section \ref{sec-discuss} we close the present article with a brief summary of all the findings.

\section{Dynamical Dark energy}
\label{sec-2}

Our universe in very large scales is almost homogeneous and isotropic. This description is  well described by a spatially flat homogeneous and isotropic geometry characterized by the Friedmann-Lema\^{i}tre-Robertson-Walker (FLRW) line element  
$ds^2 = -dt^2 + a ^2 (t) \left(dx^2 + dy^2 +dz^2\right)$, where $a(t)$ is the scale factor of the universe and  (t, x, y, z) are the comoving coordinates. The underlying theory of gravity assumed to be described by the General theory of Relativity and the distribution of matter in the universe is further assumed to be minimally coupled to the gravitational sector. Thus, within this framework, the Einstein field equations can be recast as

\begin{eqnarray}
3H^2 = \frac{8 \pi G}{3} \sum_{i} \rho_i, \label{F1}\\
2 \dot{H} + 3 H^2 = - 4 \pi G \sum_{i} p_i~,  \label{F2}
\end{eqnarray}
where $(\rho_i, p_i)$ are the energy density and pressure of the $i$-th fluid respectively, an overhead dot represents the cosmic time differentiation and 
$H \equiv \dot{a}/a$ is the Hubble rate of the FLRW universe. And we assume that none of the fluid under consideration is interacting with other fluids. Therefore, the conservation equation for each fluid follows $\dot{\rho}_i + 3 H (p_i+\rho_i) = 0$. Let us note that the conservation equation can also be derived from the above two equations, namely, equations (\ref{F1}) and (\ref{F2}).  We consider four different non-interacting components in the universe sector, namely, radiation ($i =r$), baryons ($i =b$), pressure-less dark matter ($i = c$) and a dark energy fluid ($i = x$), with their barotropic equation of state parameters as $w_r = p_r/\rho_r = 1/3$, $w_b = p_b/\rho_b =0$, $w_c  = p_c/\rho_c = 0$ and $w_x = p_x/\rho_x$, which could be any function of the scale factor/redshift. Now, using the conservation equation for each fluid, one could write down the Hubble expansion as 

\begin{eqnarray}\label{Hubble-expansion}
H^2 = \frac{8\pi G}{3} \Bigg[ \rho_{r0} a^{-4} + \rho_{b0} a^{-3} +\rho_{c0}a^{-3} + \nonumber\\+ \rho_{x,0}\,\left(  \frac{a}{a_{0}}\right)  ^{-3} \, \exp\left( -3\int_{a_{0}}^{a}\frac{w_{x} \left( a'\right) }{a'}\,da'
\right) \Bigg]
\end{eqnarray}
where $\rho_{i0}$'s ($i = r, b, c, x$) denotes the present day value of $\rho_{i}$ ($i = r, b, c, x$). Thus, whenever the dark energy equation of state $w_x (a)$ is given, one can determine the Hubble rate using eqn.  (\ref{Hubble-expansion}). In the present article we focus on the four very well known dark energy equation of state parameters given below.

\begin{itemize}

\item {\bf Parametrization 1:} We consider the simple parametrization in which the dark energy equation of state is time independent. This cosmological model is commonly known as $w$CDM cosmology.

\item {\bf Parametrization 2:} This parametrization was independently proposed by Chevallier and Polarski in {\color{blue}\cite{Chevallier:2000qy}} and later by Linder {\color{blue}\cite{Linder:2002et}} as follows:
\begin{eqnarray}    
w_x(a) = w_0 + w_a (1-a),\label{cpl}.
\end{eqnarray}

\item {\bf Parametrization 3:} Another parametrization of dark energy equation of state proposed by Jassal-Bagla-Padmanabhan {\color{blue}\cite{Jassal:2005qc}}, known by JBP parametrization which takes the following expression 
\begin{eqnarray}
w_x(a) = w_0 + w_a a (1-a),\label{jbp}.
\end{eqnarray}

\item {\bf Parametrization 4:} The dark energy equation of state allowing a logarithmic term as
\begin{eqnarray}
 w_x (a) = w_0 - w_a \ln a,\label{log}
\end{eqnarray}
was proposed by G.~Efstathiou {\color{blue}\cite{Efstathiou:1999tm}}. 
This parametrization is commonly known as Logarithmic parametrization.

\item {\bf Parametrization 5:} We introduce the last parametrization in this series proposed by Barboza and Alcaniz {\color{blue}\cite{Barboza:2008rh}} and hence it is known as BA parametrization. This parametrization has the following evolution law  
    
\begin{eqnarray}
w_x (a) = w_0 + w_a \left(\frac{1-a}{2a^2-2a+1}. \right),\label{ba}
\end{eqnarray}

\end{itemize}

For all the previous ($w_0$,$w_a$) parametrizations,  shown in eqns. (\ref{cpl}), (\ref{log}), (\ref{jbp}) and (\ref{ba}), $w_0$ is the present value of the dark energy equation of state parameter and $w_a$ gives the evolution of $w(a)$ with redshift. 
For all of them one can determine the evolution of the Hubble expansion (\ref{Hubble-expansion}). 
Let us note that for $w$CDM cosmology, henceforth of this article, we shall use $w_0$ instead of $w$ just to keep similarity with rest of the parametrizations.  

\begin{figure*}
\includegraphics[width=0.45\textwidth]{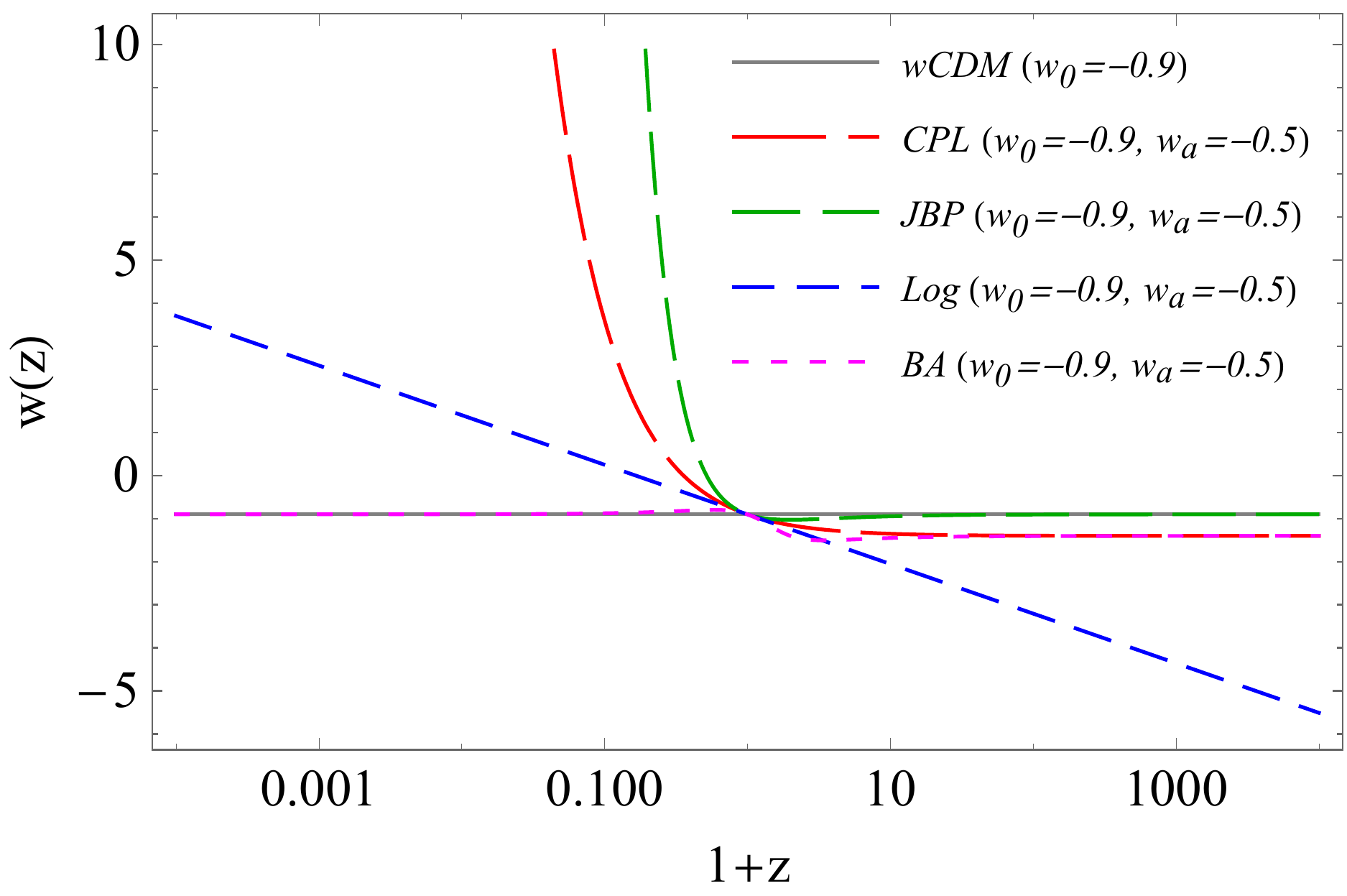}
\includegraphics[width=0.45\textwidth]{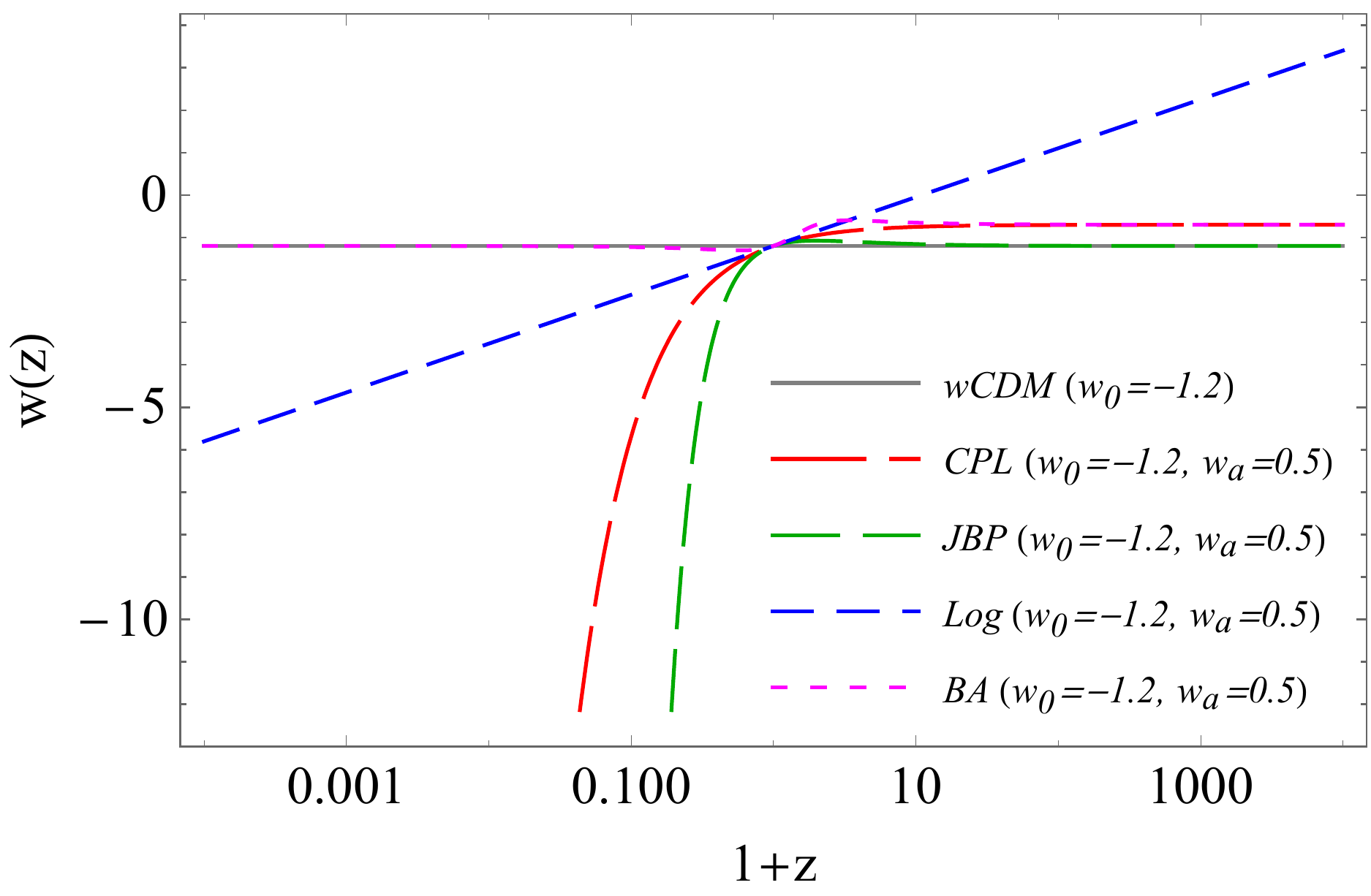}
\caption{Comparison of the behaviours of the Dark Energy equation of state $w(z) (\equiv w_x (a))$ analysed in this work, at low and high redshift. In the left panel we show the evolution of the dark energy parametrizations studied in this work where the current value of the dark energy equation of state lies in the quintessence regime $w_0>-1$, and in the right panel we consider the case when the current value of the dark energy equation of state lies in phantom regime $w_0<-1$.}
\label{fig-wz}
\end{figure*}

Lastly, in Fig.~\ref{fig-wz} we present the behaviour of all the dark energy parametrizations at low and high redshifts considering the present value of $w_0$ in both the quintessence regime (left plot of Fig.~\ref{fig-wz}) and or the phantom regime (right plot of Fig.~\ref{fig-wz}). The figure also offers a comparison between these 
parametrizations. One can see that the evolution of the CPL and JBP parametrizations are similar: they are constant at higher redshifts, and equal to $w_0+w_a$ and $w_0$ respectively, while quickly diverges towards larger (smaller) $w(z)$ values at smaller redshifts for negative (positive) $w_a$. The evolution of the BA parametrization at large redshifts is, as CPL, equal to $w_0+w_a$, but equal to $w_0$ in the low redshifts regime. 
The evolution of the Log parametrization is completely different compared to other four parametrizations. In particular, this is a straight line in a plane semi-logarithmic, crossing the point $w(z)=w_0$ when $z=0$.
Finally, we can see the effect of changing the sign of the parameter $w_a$, comparing the left and right panels of Fig.~\ref{fig-wz}.

The fact that the different models behave differently with time, is the reason why they fit in different ways the low redshift data, as the Baryon acoustic oscillations (BAO) measurements (focused in the next sections), allowing us to discriminate which is the best model to solve the tensions we have in the data.  

\section{Observational data}
\label{sec-data}

In this section, we briefly present the cosmological data sets and the statistical methodology to constrain all the dynamical dark energy models we considered in this work. Let us take into consideration:

\begin{enumerate}

\item \textbf{Cosmic Microwave Background (CMB) from Planck 2018:} We consider the latest CMB primordial power spectra measurements, in temperature and polarization, from Planck {\color{blue}\cite{Aghanim:2018eyx,Aghanim:2018oex,Aghanim:2019ame}}. We refer to this dataset at \textbf{Planck 2018}. 

\item \textbf{Cosmic Microwave Background (CMB) from Planck 2015:} Along with the latest CMB measurements from Planck 2018 we also consider its earlier temperature and polarization power spectra measurements by Planck {\color{blue}\cite{Ade:2015xua,Adam:2015rua,Aghanim:2015xee}} in order to perform a comparisons between the effects appearing from the old and new CMB measurements. We refer to this dataset at \textbf{Planck 2015}.

\item \textbf{Baryon acoustic oscillation (BAO) distance measurements}: We use various measurements of the BAO data from {\color{blue}\cite{Beutler:2011hx,Ross:2014qpa,Alam:2016hwk}}. We refer to this dataset at \textbf{BAO}.

\item \textbf{Hubble Space Telescope (HST):}  We also consider two estimates of the Hubble constant obtained by analysing the Hubble Space Telescope measurements, using Cepheids as geometric distance calibrators. They report $H_0 = 73.48 \pm 1.66$ at 68\% CL {\color{blue}\cite{Riess:2018uxu}} and 
$H_0 = 74.03 \pm 1.42$ km/s/Mpc at $68\%$ CL~{\color{blue}\cite{Riess:2019cxk}}. Let us note that the estimations of $H_0$ from {\color{blue}\cite{Riess:2018uxu,Riess:2019cxk}} are in huge tension ($> 4 \sigma$) with the Planck's 2015 and 2018 measurements within the minimal $\Lambda$CDM scenario. We refer to the $H_0$ measurement from \cite{Riess:2018uxu} as \textbf{R18} and from {\color{blue}\cite{Riess:2019cxk}} as \textbf{R19}.

\end{enumerate}

To constrain the underlying cosmological scenarios driven by the above dark energy parametrizations, we modified the Markov Chain Monte Carlo package \texttt{CosmoMC} {\color{blue}\cite{Lewis:2002ah,Lewis:2013hha}}. \texttt{CosmoMC} is an excellent cosmological code used extensively to extract the cosmological parameters constraints analysing the data. This code  is also equipped with a convergence diagnostic by Gelman and Rubin {\color{blue}\cite{Gelman-Rubin}}. Additionally, the \texttt{CosmoMC} package also supports the Planck 2015 and Planck 2018 likelihoods. 

As a baseline model we consider the six parameters of the $\Lambda$CDM model: baryon and cold dark matter density $\Omega_{\rm b} h^2$ and $\Omega_{\rm c} h^2$, the optical depth to reionization $\tau$, the amplitude and the spectral index of the primordial power spectrum of scalar perturbations $n_s$ and the angular size of the sound horizon at decoupling $\theta_{MC}$. Moreover we add the parameters of the dark energy models we are considering: $w_0$ in $w$CDM or ($w_0,w_a$) in all the other cases. In Table \ref{tab:priors} we display the flat priors adopted on the cosmological parameters for all the models considered during the statistical analyses.

\begin{table}[ht]
\begin{center}
\begin{tabular}{|c|c|}
\hline
Parameter                    & Prior\\
\hline
$\Omega_{\rm b} h^2$         & $[0.005,0.1]$\\
$\Omega_{\rm c} h^2$         & $[0.01, 0.99]$\\
$\tau$                       & $[0.01,0.8]$\\
$n_s$                        & $[0.5, 1.5]$\\
$\log[10^{10}A_{s}]$         & $[2.4,4]$\\
$100\theta_{MC}$             & $[0.5,10]$\\
$w_0$                        & $[-2, 0]$\\
$w_a$                        & $[-3, 3]$\\
\hline
\end{tabular}
\end{center}
\caption{Flat priors on various free parameters of all the models used in this work. For the Logarithmic model the $w_a$ range is instead $[-3, 0]$ for stability reasons.}
\label{tab:priors}
\end{table}

\section{Results and analysis}
\label{sec-results}

\begin{figure*}
\includegraphics[width=0.6\textwidth]{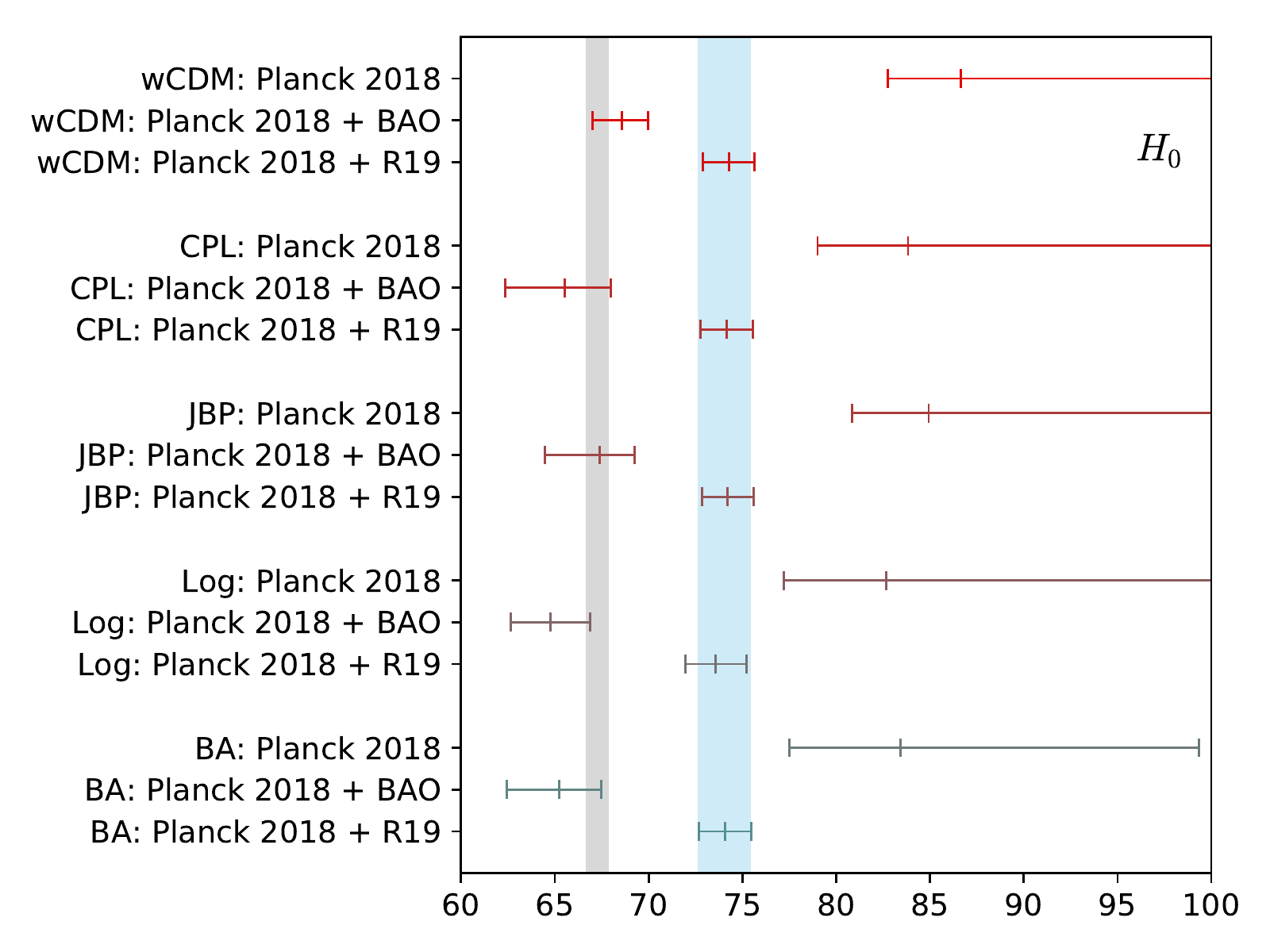}
\caption{We show the whisker plot of the Hubble constant $H_0$ considering its 68\%~CL constraints obtained for cosmological analysis of dynamical dark energy with Planck 2018, BAO and R19. The grey vertical bands correspond to the 68\%~CL Planck 2018 bound on the Hubble constant under the assumption of a $\Lambda$CDM model, while the cyan one to the R19 value measured by SH0ES in~{\color{blue}\cite{Riess:2019cxk}} at 68\%~CL.}
\label{fig2}
\end{figure*}

In this section we shall describe the main observational results that we have extracted from all the cosmological models through the use of various observational data. We have mainly focused on the estimations of $H_0$ from the underlying cosmological models driven by various dark energy parametrizations. We refer to the whisker plots for $H_0$ displayed in Fig.~\ref{fig2} (Planck 2018 and its combination with BAO and R19) for a better understanding on the estimations of $H_0$ for different datasets and models. 
In the following we have described the results for each model.

\subsection{$w$CDM parametrization}
\label{sec-wcdm}

\begin{figure*}
    \centering
    \includegraphics[width=0.65\textwidth]{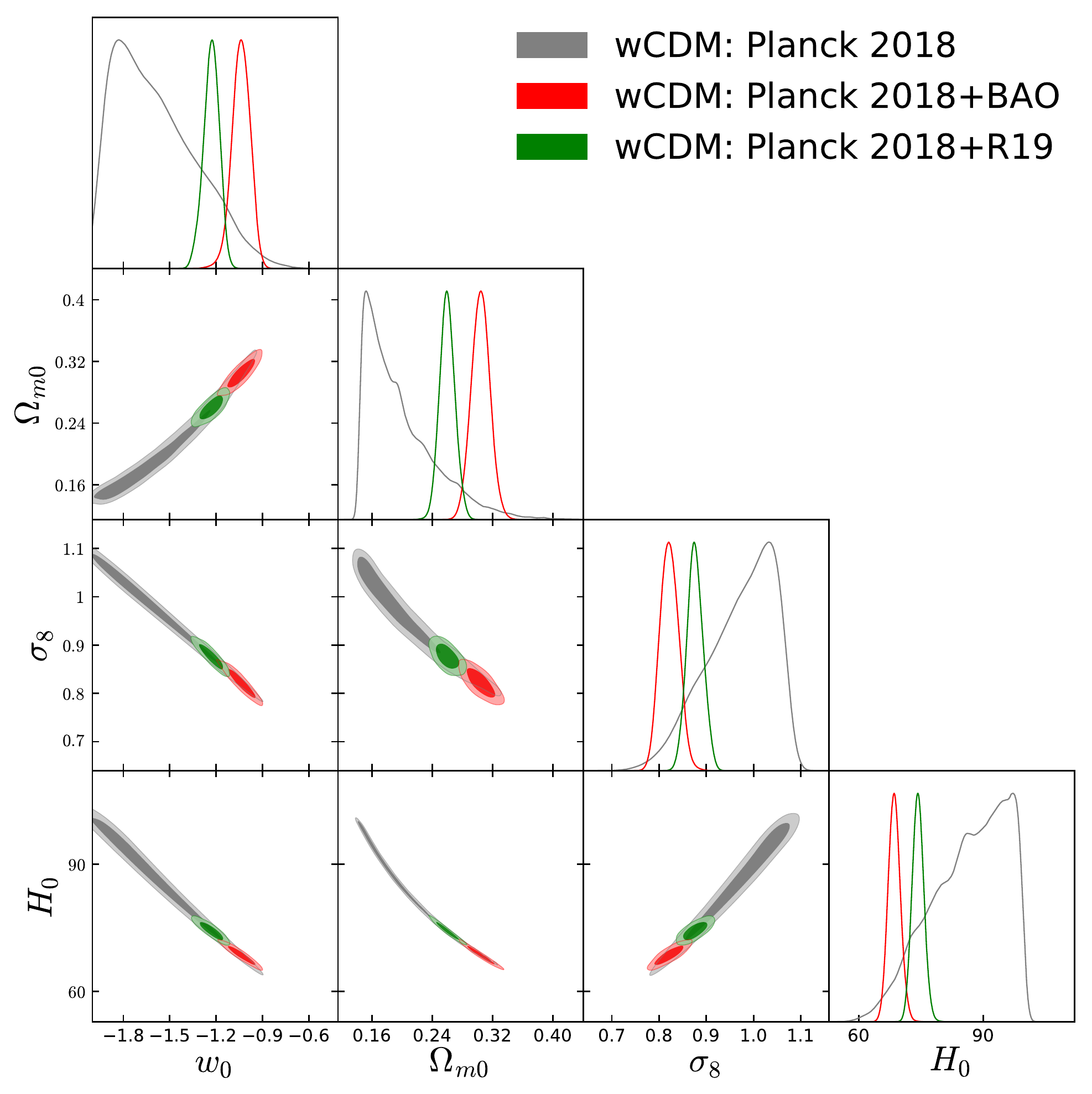}
    \caption{We show  the 1-dimensional marginalized posterior distributions and 2-dimensional contour plots of some key parameters of the $w$CDM cosmological model using Planck 2018 and its combination with BAO and R19. }
    \label{fig:tri-wcdm}
\end{figure*}

In Table \ref{tab:wCDM} we have summarized the observational constraints on this cosmological model using various cosmological datasets. In the left side of Table \ref{tab:wCDM} we have shown the constraints for Planck 2015 and its combination with BAO and R18. We note that the BAO data are added to Planck 2015 in order to break the degeneracies among the parameters. On the other hand, although Planck 2015 and R18 are in tension within the minimal $\Lambda$CDM scenario, but in this case, the estimations of $H_0$
for Planck 2015 and R18 are not in tension, so one can safely add R18 with Planck 2015. Similarly in the right panel of Table \ref{tab:wCDM}, we have shown the constraints for Planck 2018, Planck 2018+BAO and Planck 2018+R19. From our analyses we did not find any tension in $H_0$ for Planck 2018 and R19 under this model consideration, and thus, the combination Planck 2018+R19 is safe. 
In Fig. \ref{fig:tri-wcdm} we display the triangular plot for Planck 2018 and its combination with BAO and R19.  

Let us now focus on the observational constraints extracted from both Planck 2015 and Planck 2018 datasets and their combinations. 
For Planck datasets only, namely Planck 2015 and Planck 2018, we find that $H_0$ attains very large values with significantly large error bars. This is due to the geometrical degeneracy present with $w_0$, that assumes values less than $-1$ and remains in the phantom regime at more than 68\% CL for Planck 2015 and more than 95\% for Planck 2018. We can see their negative correlation in Fig. \ref{fig:tri-wcdm}. 
Therefore, going from Planck 2015 to Planck 2018 we see a shift of the mean value of $H_0$ towards higher values, corresponding to the shift towards a more phantom $w$ behaviour in the Planck 2018 data. This is probably due to the lower and better constrained optical depth $\tau$ present in the latest Planck release.

When BAO data are added to Planck 2015 and Planck 2018, we find significant improvements in constraining the $H_0$ parameter, but shifting its estimates towards lower values. Hence, for Planck 2015+BAO and Planck 2018+BAO, the tension on $H_0$ is restored, however, due to the slightly larger error bars than in the $\Lambda$CDM case, the $H_0$ tension is alleviated at about $2.8\sigma$ and $2.7\sigma$, respectively. For these dataset combinations the cosmological constant $w=-1$ is back in agreement with the data within one standard deviation.

Finally, when we consider Planck 2015+R18 and Planck 2018+R19, $H_0$, that was almost unconstrained in the Planck alone case, is now fixed by the R18 and R19 measurement respectively. Moreover, the addition of the Hubble constant priors produces very important bounds on $w$ that is here more different from the cosmological constant than in the Planck alone cases.

\begingroup
\squeezetable
\begin{center}
\begin{table*}
\scalebox{0.9}{
\begin{tabular}{c|ccc|ccc}
\hline\hline
Parameters & Planck 2015 & Planck 2015+BAO & Planck 2015+R18 & Planck 2018 & Planck 2018+BAO & Planck 2018+R19 \\ \hline
$\Omega_ch^2$&
$0.1192_{-0.0014-0.0027}^{+0.0014+0.0028}$&
$0.1185_{-0.0014-0.0024}^{+0.0013+0.0025}$&
$0.1191_{-0.0014-0.0029}^{+0.0014+0.0027}$&
$0.1199_{-0.0013-0.0026}^{+0.0013+0.0027}$&
$0.1199_{-0.0013-0.0026}^{+0.0013+0.0025}$&
$0.1202_{-0.0013-0.0026}^{+0.0013+0.0025}$
\\
$\Omega_bh^2$&
$0.02227_{-0.00015-0.00029}^{+0.00015+0.00030}$&
$0.02230_{-0.00015-0.00030}^{+0.00016+0.00029}$&
$0.02227_{-0.00015-0.00030}^{+0.00015+0.00029}$&
$0.02240_{-0.00015-0.00030}^{+0.00015+0.00030}$&
$0.02238_{-0.00015-0.00028}^{+0.00015+0.00029}$&
$0.02237_{-0.00017-0.00028}^{+0.00014+0.00031}$
\\
$100\theta_{MC}$&
$1.04079_{-0.00034-0.00062}^{+0.00032+0.00064}$&
$1.04087_{-0.00032-0.00062}^{+0.00032+0.00062}$&
$1.04077_{-0.00031-0.00061}^{+0.00032+0.00066}$&
$1.04093_{-0.00031-0.00060}^{+0.00031+0.00062}$&
$1.04095_{-0.00031-0.00062}^{+0.00030+0.00060}$&
$1.04091_{-0.00032-0.00063}^{+0.00033+0.00063}$
\\
$\tau$&
$0.078_{-0.017-0.036}^{+0.017+0.034}$&
$0.084_{-0.017-0.034}^{+0.018+0.033}$&
$0.079_{-0.017-0.033}^{+0.019+0.033}$&
$0.0543_{-0.0082-0.015}^{+0.0074+0.016}$&
$0.0547_{-0.0076-0.015}^{+0.0077+0.015}$&
$0.0541_{-0.0079-0.015}^{+0.0069+0.017}$
\\
$n_s$&
$0.9664_{-0.0047-0.0089}^{+0.0046+0.0089}$&
$0.9678_{-0.0044-0.0083}^{+0.0045+0.0090}$&
$0.9665_{-0.0050-0.0086}^{+0.0043+0.0096}$&
$0.9654_{-0.0043-0.0087}^{+0.0043+0.0087}$&
$0.9656_{-0.0043-0.0080}^{+0.0041+0.0087}$&
$0.9648_{-0.0041-0.0081}^{+0.0040+0.0083}$
\\
${\rm{ln}}(10^{10}A_s)$&
$3.089_{-0.034-0.064}^{+0.034+0.065}$&
$3.100_{-0.034-0.066}^{+0.034+0.065}$&
$3.090_{-0.035-0.066}^{+0.034+0.063}$&
$3.044_{-0.017-0.031}^{+0.015+0.033}$&
$3.045_{-0.015-0.032}^{+0.015+0.031}$&
$3.044_{-0.017-0.030}^{+0.015+0.034}$
\\
$w_0$&
$-1.47_{-0.43}^{+0.23} <-0.91$&
$-1.021_{-0.052-0.11}^{+0.056+0.11}$&
$-1.195_{-0.056-0.12}^{+0.058+0.12}$&
$-1.59_{-0.36}^{+0.15} <-1.11$&
$-1.039_{-0.055-0.12}^{+0.060+0.12}$&
$-1.231_{-0.049-0.11}^{+0.051+0.10}$
\\
$\Omega_{m0}$&
$0.216_{-0.077-0.095}^{+0.032+0.122}$&
$0.303_{-0.013-0.022}^{+0.011+0.024}$&
$0.263_{-0.014-0.024}^{+0.011+0.026}$&
$0.197_{-0.058-0.069}^{+0.020+0.098}$&
$0.304_{-0.012-0.024}^{+0.012+0.025}$&
$0.260_{-0.010-0.019}^{+0.010+0.021}$
\\
$\sigma_8$&
$0.96_{-0.07-0.17}^{+0.12+0.15}$&
$0.836_{-0.021-0.041}^{+0.020+0.039}$&
$0.884_{-0.022-0.042}^{+0.021+0.041}$&
$0.973_{-0.047-0.15}^{+0.097+0.12}$&
$0.822_{-0.020-0.038}^{+0.020+0.039}$&
$0.876_{-0.018-0.033}^{+0.017+0.034}$
\\
$H_0\,{\rm [km/s/Mpc]}$&
$83_{-9}^{+14} >64$&
$68.4_{-1.4-2.8}^{+1.4+2.7}$&
$73.6_{-1.6-3.5}^{+1.8+3.2}$&
$>81 >69$&
$68.6_{-1.6-2.8}^{+1.4+3.0}$&
$74.3_{-1.4-2.7}^{+1.4+2.9}$
\\
\hline\hline
\end{tabular}
}
\caption{Observational constraints on free and derived parameters at 68\% and 95\% CL in the $w$CDM cosmological model for various observational data and their combinations. }
\label{tab:wCDM}
\end{table*}
\end{center}
\endgroup

\subsection{CPL parametrization}
\label{sec-cpl}

\begin{figure*}
    \centering
    \includegraphics[width=0.65\textwidth]{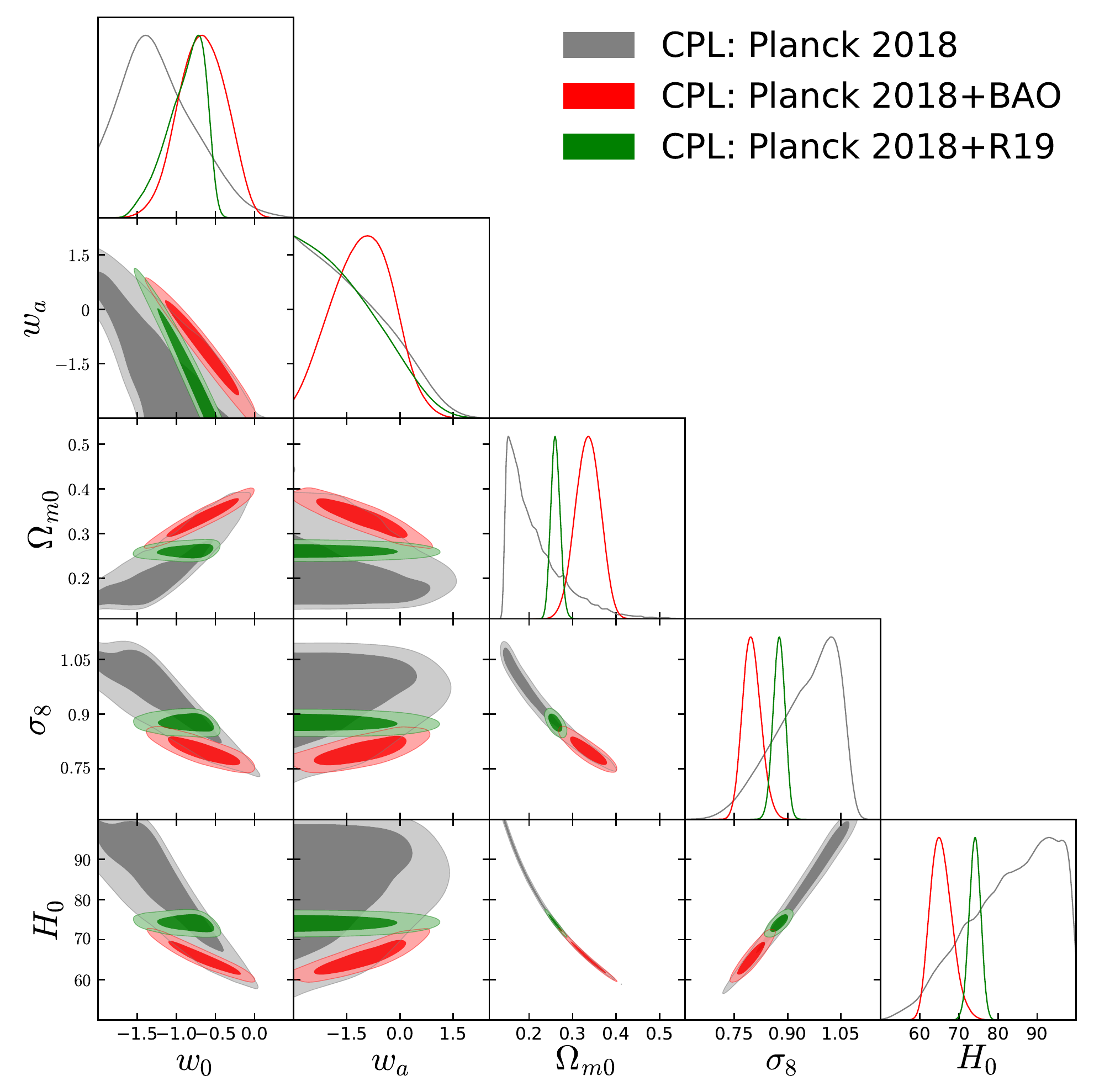}
    \caption{We show the 1-dimensional marginalized posterior distributions and 2-dimensional contour plots of some key parameters of the CPL cosmological model using Planck 2018 and its combination with BAO and R19. }
    \label{fig:tri-cpl}
\end{figure*}

We now consider the CPL parametrization of eqn. (\ref{cpl}) the results of which are summarized in Table \ref{tab:CPL}.  In a similar way we have considered the CMB data from both Planck 2015 and Planck 2018 missions and considered other statistical analyses using BAO and local measurements of $H_0$. 
The left panel of Table \ref{tab:CPL} refers to   
the constraints from the model for Planck 2015 data, Planck 2015+BAO and Planck 2015+R18. Similarly, the right panel of~\ref{tab:CPL} refers to the constraints from the model for Planck 2018 data, Planck 2018+BAO and Planck 2018+R19. 
As already commented in subsection~\ref{sec-wcdm}, the combination of CMB data from Planck and $H_0$ data from local measurements is safe because the tension between them is relieved in the CPL model. Thus, we can consider their combinations, namely,  Planck 2015+R18 and Planck 2018+R19. Similarly, to the previous section, in Fig.~\ref{fig:tri-cpl} we display the triangular plot for Planck 2018 and its combination with BAO and R19.

Also in this case, we find that for CPL model, the estimations of $H_0$ for both Planck 2015 and Planck 2018 have very high mean values together with very large error bars.  
This is due to the negative correlation there is between $H_0$ and $w_0$, that prefer a mean value less than $-1$, while there is not significant correlation between $H_0$ and $w_a$ \ref{fig:tri-cpl}. 
Moreover, the $H_0$ and $w$ constraints are really robust going from Planck 2015 to Planck 2018, so we don't appreciate any significant difference.

When we add BAO data to the Planck datasets, namely, Planck 2015 and Planck 2018, we find that the mean value of $H_0$ is much lower compared to Planck's alone estimations. As one can see the 68\% CL (and 95\% CL) constraints on $H_0$ are 
$H_0 = 65.4_{-2.9}^{+2.4}$ km/s/Mpc ($H_0= 65.4_{-5.0}^{+5.1}$ km/s/Mpc) for Planck 2015+BAO, and $H_0 = 65.5_{-3.2}^{+2.4}$ km/s/Mpc ($H_0= 65.5_{-5.3}^{+5.7}$ km/s/Mpc) for Planck 2018+BAO. Even if we have very large error bars on $H_0$, the tension with R19 is restored at about $3\sigma$. Looking at the dark energy equation of state, a quintessence value $w_0>-1$ and $w_a<0$ are both preferred at more than one standard deviation. Moreover, the addition of the BAO data introduces a correlation between $H_0$ and $w_a$, as we can see in Fig.~\ref{fig:tri-cpl}, that was not present for the Planck alone dataset.

Finally, when we consider the final two datasets, namely, Planck 2015+R18 and Planck 2018+R19, since $H_0$ was almost unconstrained in the Planck alone cases, is now completely fixed by the R18 and R19 measurements respectively. As one can see, $H_0 = 73.5_{-1.6-3.3}^{+1.6+3.2}$ km/s/Mpc (Planck 2015+R18 at 68\% CL) and 
$H_0 =  74.1_{-1.4-2.8}^{+1.4+2.8}$ km/s/Mpc (Planck 2018+R19, at 68\% CL). 
Moreover, the addition of the Hubble constant priors produces a shift of the $w_0$ constraint more in agreement with the Planck + BAO results, while $w_a$ remains almost unconstrained like in the Planck alone case, as we can see in Fig.~\ref{fig:tri-cpl}.

\begingroup
\squeezetable
\begin{center}
\begin{table*}
\scalebox{0.9}{
\begin{tabular}{c|ccc|ccc}
\hline\hline
Parameters & Planck 2015 & Planck 2015+BAO & Planck 2015+R18 & Planck 2018 & Planck 2018+BAO & Planck 2018+R19 \\ \hline
$\Omega_ch^2$&
$0.1190_{-0.0014-0.0027}^{+0.0014+0.0028}$&
$0.1193_{-0.0014-0.0026}^{+0.0013+0.0026}$&
$0.1191_{-0.0014-0.0027}^{+0.0014+0.0028}$&
$0.1198_{-0.0014-0.0027}^{+0.0014+0.0027}$&
$0.1201_{-0.0013-0.0024}^{+0.0012+0.0025}$&
$0.1199_{-0.0014-0.0027}^{+0.0014+0.0027}$
\\
$\Omega_bh^2$&
$0.02228_{-0.00015-0.00030}^{+0.00016+0.00031}$&
$0.02225_{-0.00015-0.00028}^{+0.00015+0.00031}$&
$0.02227_{-0.00015-0.00031}^{+0.00016+0.00031}$&
$0.02240_{-0.00015-0.00029}^{+0.00015+0.00030}$&
$0.02237_{-0.00015-0.00028}^{+0.00014+0.00028}$&
$0.02239_{-0.00015-0.00029}^{+0.00015+0.00029}$
\\
$100\theta_{MC}$&
$1.04080_{-0.00032-0.00063}^{+0.00032+0.00064}$&
$1.04077_{-0.00032-0.00062}^{+0.00032+0.00064}$&
$1.04078_{-0.00032-0.00065}^{+0.00033+0.00065}$&
$1.04098_{-0.00031-0.00064}^{+0.00032+0.00062}$&
$1.04093_{-0.00031-0.00059}^{+0.00030+0.00060}$&
$1.04093_{-0.00031-0.00061}^{+0.00031+0.00060}$
\\
$\tau$&
$0.075_{-0.018-0.034}^{+0.017+0.034}$&
$0.079_{-0.017-0.034}^{+0.017+0.032}$&
$0.077_{-0.017-0.034}^{+0.018+0.034}$&
$0.0541_{-0.0080-0.015}^{+0.0074+0.016}$&
$0.0539_{-0.0085-0.015}^{+0.0073+0.016}$&
$0.0538_{-0.0074-0.015}^{+0.0076+0.015}$
\\
$n_s$&
$0.9665_{-0.0046-0.0090}^{+0.0046+0.0090}$&
$0.9660_{-0.0044-0.0082}^{+0.0046+0.0081}$&
$0.9664_{-0.0046-0.0090}^{+0.0045+0.0091}$&
$0.9657_{-0.0044-0.0086}^{+0.0044+0.0087}$&
$0.9651_{-0.0041-0.0084}^{+0.0041+0.0083}$&
$0.9653_{-0.0044-0.0085}^{+0.0043+0.0085}$
\\
${\rm{ln}}(10^{10}A_s)$&
$3.083_{-0.034-0.066}^{+0.034+0.066}$&
$3.090_{-0.032-0.065}^{+0.033+0.061}$&
$3.087_{-0.034-0.068}^{+0.034+0.066}$&
$3.044_{-0.015-0.031}^{+0.016+0.032}$&
$3.044_{-0.017-0.031}^{+0.015+0.034}$&
$3.043_{-0.015-0.031}^{+0.015+0.031}$
\\
$w_0$&
$-1.18_{-0.56-0.77}^{+0.32+0.87}$&
$-0.68_{-0.30-0.50}^{+0.26+0.53}$&
$-0.90_{-0.19-0.56}^{+0.36+0.46}$&
$-1.21_{-0.60}^{+0.33}  <-0.37$&
$-0.67_{-0.32-0.56}^{+0.32+0.61}$&
$-0.89_{-0.16-0.51}^{+0.32+0.41}$
\\
$w_a$&
$<-0.76 <0.43$&
$-0.96_{-0.67-1.5}^{+0.89+1.3}$&
$<-0.80<0.72$&
$<-0.85<0.71$&
$-1.05_{-0.77-1.7}^{+0.99+1.5}$&
$<-1.04<0.47$
\\
$\Omega_{m0}$&
$0.219_{-0.082-0.09}^{+0.027+0.14}$&
$0.334_{-0.026-0.049}^{+0.027+0.050}$&
$0.263_{-0.013-0.024}^{+0.011+0.025}$&
$0.215_{-0.078-0.09}^{+0.024+0.14}$&
$0.335_{-0.029-0.055}^{+0.029+0.056}$&
$0.261_{-0.011-0.020}^{+0.010+0.021}$
\\
$\sigma_8$&
$0.96_{-0.06-0.18}^{+0.12+0.15}$&
$0.813_{-0.029-0.052}^{+0.025+0.053}$&
$0.886_{-0.020-0.039}^{+0.020+0.039}$&
$0.95_{-0.06-0.18}^{+0.12+0.14}$&
$0.799_{-0.029-0.050}^{+0.023+0.053}$&
$0.876_{-0.017-0.033}^{+0.017+0.031}$
\\
$H_0\,{\rm [km/s/Mpc]}$&
$83_{-8}^{+16} >62$&
$65.4_{-2.9-5.0}^{+2.4+5.1}$&
$73.5_{-1.6-3.3}^{+1.6+3.2}$&
$84_{-7}^{+15} >63$&
$65.5_{-3.2-5.3}^{+2.4+5.7}$&
$74.1_{-1.4-2.8}^{+1.4+2.8}$
\\
\hline\hline
\end{tabular}
}
\caption{Observational constraints on various free and derived parameters at 68\% and 95\% CL in the cosmological model where DE obeys the CPL parametrization, for various observational data and their combinations. }
\label{tab:CPL}
\end{table*}
\end{center}
\endgroup

\subsection{JBP parametrization}
\label{sec-jbp}

\begin{figure*}
    \centering
    \includegraphics[width=0.65\textwidth]{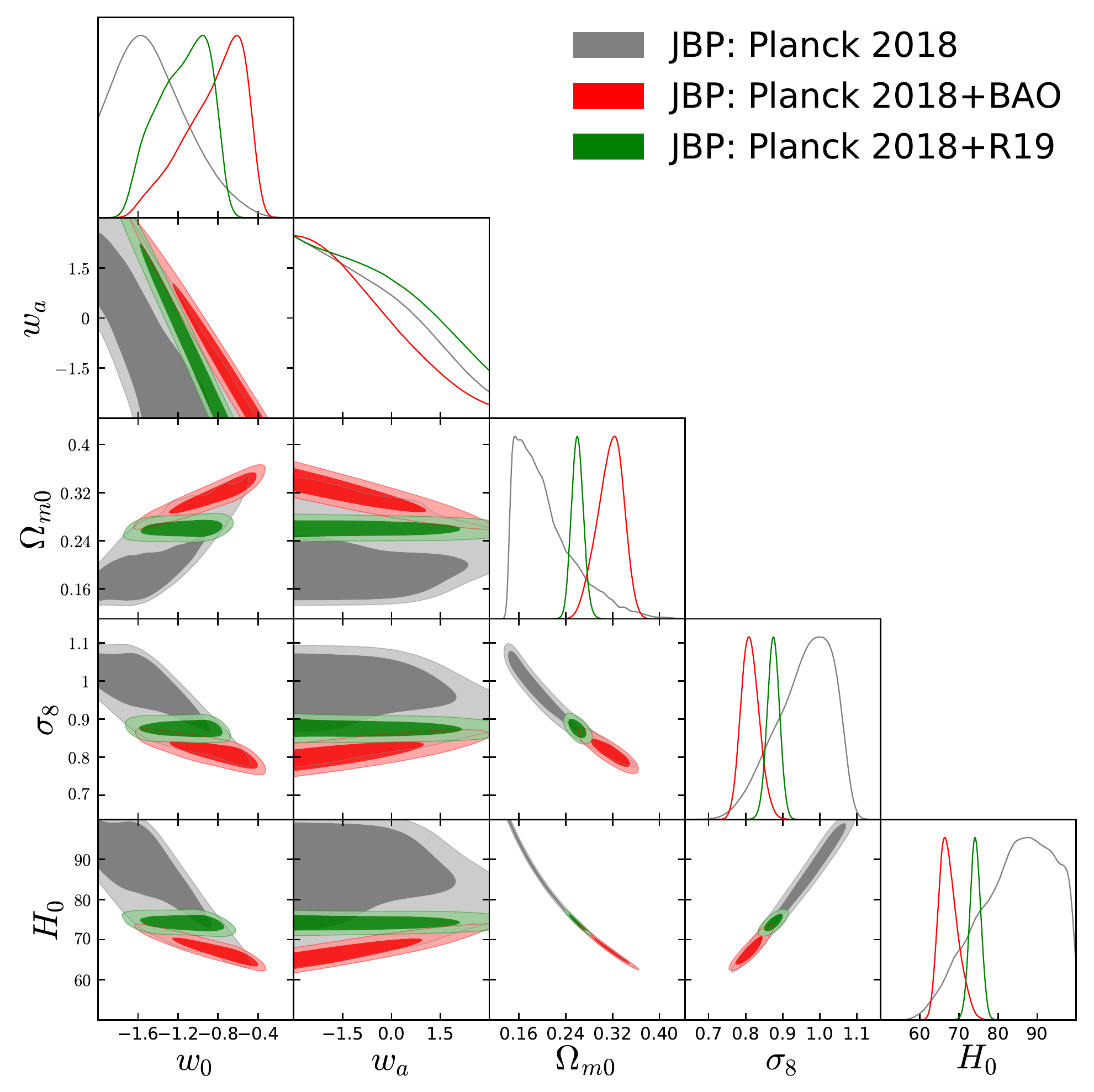}
    \caption{We show the 1-dimensional marginalized posterior distributions and 2-dimensional contour plots of some key parameters of the JBP cosmological model using Planck 2018 and its combination with BAO and R19.}
     \label{fig:tri-jbp}
\end{figure*}

In Table~\ref{tab:JBP} we have shown the results for the JBP DE parametrization using various observational data. Similarly to the previous models, here too, we have considered six different observational data combinations with CMB from Planck 2015 and Planck 2018, BAO, R18 and R19. Again, in Table~\ref{tab:JBP} we have on the left the constraints on the model from Planck 2015 data, Planck 2015+BAO and Planck 2015+R18. On the right, instead, we have the constraints on the model from Planck 2018 data, Planck 2018+BAO and Planck 2018+R19. 
Here too we can safely combine Planck 2015+R18 and Planck 2018+R19 because for the JBP model the tension is solved. Finally, in Fig. \ref{fig:tri-jbp} we display the triangular plot for Planck 2018 and its combination with BAO and R19.

Similarly to the previous two models, we see that CMB alone, namely Planck 2015 and Planck 2018, gives $H_0$ almost unconstrained. This is due to the well known geometrical degeneracy between $H_0$ and the dark energy equation of state parameter $w_0$. In this JBP parametrization we find $w_0<-1$ at more than 68\% CL
and $w_a$ unconstrained at 95\% CL, for Planck 2015, while the Planck 2018 data give a better constrained $w_a$, showing an upper limit at 95\% CL too.

When BAO data are added to Planck, we find that $H_0$ is significantly lowered with respect to the Planck alone cases, and consistent with its estimations in a $\Lambda$CDM scenario but with larger error bars. This means that the inclusion of BAO data restores again the $H_0$ tension, that is now at about $2.8\sigma$ for both Planck 2015+BAO and Planck 2018+BAO. As a consequence, $w_0$ is again completely consistent with a cosmological constant within one standard deviation, but $w_a$ is still negative at more than 68\% CL.

Finally, we consider the addition of the local measurements of $H_0$ to Planck. We see that for 
both the datasets, namely Planck 2015+R18 and Planck 2018+R19, because of the very large error bars on $H_0$ in the Planck alone cases, the Hubble constant is now completely determined by the R18 and R19 priors respectively. As one can see, $H_0  = 73.6\pm1.6$ km/s/Mpc (68\% CL for Planck 2015+R18) and 
$H_0 = 74.2\pm1.4$ km/s/Mpc (68\% CL, Planck 2018+R19).
Moreover, also in this case, the addition of the Hubble constant priors breaks completely the degeneracy between the Hubble parameter and the dark energy equation of state parameters, as we can see in Fig.~\ref{fig:tri-jbp}. The dark energy equation of state parameters ($w_0$,$w_a$) are in agreement with a cosmological constant ($w_0=-1$,$w_a=0$) within $1\sigma$ for both Planck 2015+R18 and Planck 2018+R19.

\begingroup
\squeezetable
\begin{center}
\begin{table*}
\scalebox{0.9}{
\begin{tabular}{c|ccc|ccc}
\hline\hline
Parameters & Planck 2015 & Planck 2015+BAO & Planck 2015+R18 & Planck 2018 & Planck 2018+BAO & Planck 2018+R19 \\ \hline
$\Omega_ch^2$&
$0.1191_{-0.0014-0.0028}^{+0.0014+0.0029}$&
$0.1191_{-0.0013-0.0027}^{+0.0013+0.0025}$&
$0.1192_{-0.0014-0.0027}^{+0.0014+0.0027}$&
$0.1199_{-0.0014-0.0026}^{+0.0014+0.0027}$&
$0.1199_{-0.0012-0.0025}^{+0.0013+0.0024}$&
$0.1201_{-0.0013-0.0027}^{+0.0013+0.0026}$
\\
$\Omega_bh^2$&
$0.02228_{-0.00016-0.00031}^{+0.00015+0.00031}$&
$0.02227_{-0.00015-0.00028}^{+0.00015+0.00029}$&
$0.02226_{-0.00015-0.00030}^{+0.00016+0.00031}$&
$0.02240_{-0.00015-0.00029}^{+0.00015+0.00029}$&
$0.02238_{-0.00014-0.00029}^{+0.00014+0.00028}$&
$0.02238_{-0.00016-0.00029}^{+0.00014+0.00030}$
\\
$100\theta_{MC}$&
$1.04080_{-0.00033-0.00066}^{+0.00035+0.00064}$&
$1.04081_{-0.00031-0.00061}^{+0.00031+0.00061}$&
$1.04077_{-0.00032-0.00064}^{+0.00032+0.00064}$&
$1.04093_{-0.00031-0.00061}^{+0.00031+0.00061}$&
$1.04094_{-0.00029-0.00059}^{+0.00029+0.00059}$&
$1.04093_{-0.00030-0.00061}^{+0.00030+0.00058}$
\\
$\tau$&
$0.076_{-0.018-0.035}^{+0.018+0.034}$&
$0.082_{-0.017-0.033}^{+0.017+0.033}$&
$0.078_{-0.017-0.034}^{+0.017+0.034}$&
$0.0539_{-0.0083-0.015}^{+0.0073+0.016}$&
$0.0544_{-0.0080-0.015}^{+0.0073+0.016}$&
$0.0538_{-0.0077-0.015}^{+0.0077+0.016}$
\\
$n_s$&
$0.9664_{-0.0046-0.0091}^{+0.0045+0.0090}$&
$0.9666_{-0.0042-0.0080}^{+0.0041+0.0082}$&
$0.9661_{-0.0045-0.0090}^{+0.0045+0.0090}$&
$0.9655_{-0.0042-0.0085}^{+0.0043+0.0084}$&
$0.9654_{-0.0041-0.0080}^{+0.0043+0.0083}$&
$0.9648_{-0.0043-0.0084}^{+0.0043+0.0085}$
\\
${\rm{ln}}(10^{10}A_s)$&
$3.084_{-0.035-0.068}^{+0.036+0.066}$&
$3.095_{-0.034-0.066}^{+0.034+0.065}$&
$3.089_{-0.034-0.066}^{+0.033+0.065}$&
$3.043_{-0.016-0.030}^{+0.016+0.033}$&
$3.044_{-0.017-0.030}^{+0.015+0.032}$&
$3.044_{-0.016-0.031}^{+0.016+0.032}$
\\
$w_0$&
$-1.42_{-0.49}^{+0.22}<-0.75$&
$-0.87_{-0.24-0.57}^{+0.38+0.53}$&
$-1.13_{-0.21-0.51}^{+0.37+0.44}$&
$-1.44_{-0.47}^{+0.22}<-0.79$&
$-0.84_{-0.19-0.63}^{+0.40+0.50}$&
$-1.15_{-0.21-0.50}^{+0.35+0.43}$
\\
$w_a$&
$< 0.19, \; \mbox{unconst. }$&
$<-0.10 <1.73$&
$<0.41, \; \mbox{unconst.}$&
$<0.07<2.09$&
$<-0.42 <1.75$&
$<0.33, \; \mbox{unconst.}$
\\
$\Omega_{m0}$&
$0.210_{-0.069-0.09}^{+0.027+0.12}$&
$0.315_{-0.022-0.043}^{+0.022+0.043}$&
$0.263_{-0.013-0.022}^{+0.011+0.025}$&
$0.206_{-0.064-0.08}^{+0.022+0.11}$&
$0.316_{-0.020-0.046}^{+0.026+0.041}$&
$0.260_{-0.010-0.019}^{+0.010+0.021}$
\\
$\sigma_8$&
$0.97_{-0.06-0.16}^{+0.11+0.14}$&
$0.827_{-0.029-0.048}^{+0.025+0.053}$&
$0.885_{-0.020-0.039}^{+0.021+0.041}$&
$0.96_{-0.053-0.15}^{+0.10+0.13}$&
$0.813_{-0.028-0.048}^{+0.022+0.051}$&
$0.875_{-0.016-0.033}^{+0.017+0.032}$
\\
$H_0\,{\rm [km/s/Mpc]}$&
$84_{-8}^{+13} >65$&
$67.2_{-2.7-4.6}^{+2.1+4.8}$&
$73.6_{-1.6-3.2}^{+1.6+3.2}$&
$85_{-7}^{+13} >67$&
$67.4_{-2.9-4.4}^{+1.9+5.2}$&
$74.2_{-1.4-2.7}^{+1.4+2.8}$
\\
\hline\hline
\end{tabular}
}
\caption{Observational constraints on various free and derived parameters at 68\% and 95\% CL in the cosmological model where DE follows the JBP parametrization, for various observational data and their combinations. }
\label{tab:JBP}
\end{table*}
\end{center}
\endgroup

\subsection{Logarithmic parametrization}
\label{sec-log}

\begin{figure*}
    \centering
    \includegraphics[width=0.65\textwidth]{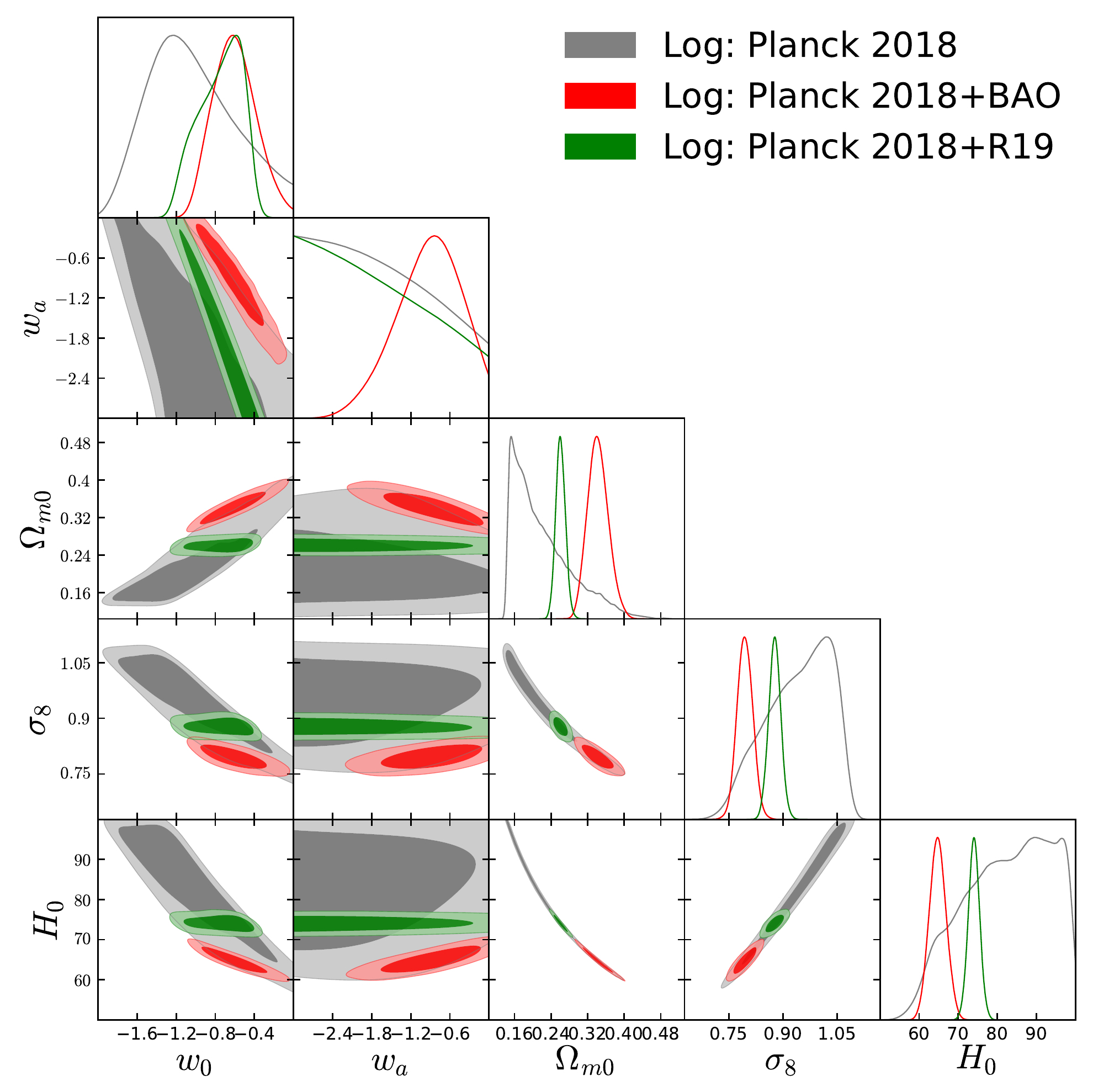}
    \caption{We show the 1-dimensional marginalized posterior distributions and 2-dimensional contour plots of some key parameters of the Logarithmic cosmological model using Planck 2018 and its combination with BAO and R19.}
     \label{fig:tri-log}
\end{figure*}

In Table \ref{tab:Log} we have summarized the observational constraints for this model using various observational data. Again, also for this model, we have considered Planck 2015 and Planck 2018 CMB data, and their combinations with other observational probes, such as BAO and local measurements of $H_0$.  
Similarly to the previous cases, in Table \ref{tab:Log} we show on the first 3 columns the constraints on the model from Planck 2015 data, Planck 2015+BAO and Planck 2015+R18. On the 4th to 6th columns of~\ref{tab:Log} we show, instead, the constraints on the model from Planck 2018 data, Planck 2018+BAO and Planck 2018+R19. 
In the Logarithmic parametrization as well, we can safely combine Planck 2015+R18 and Planck 2018+R19 because they are no more in tension. Finally,
in Fig. \ref{fig:tri-log} there is the triangular plot for Planck 2018 and its combination with BAO and R19.

Also in this case, we have for Planck alone 2015 and 2018, the estimations of $H_0$ with very large error bars at 68\% CL and just lower limits at 95\% CL. However, on the contrary with respect to the previous models, in this case $w_0$ is completely in agreement within $1\sigma$ with $-1$ and $w_a$ is unconstrained in the range $[-3;0]$. While a negative correlation between $H_0$ and $w_0$ is present for Planck alone, $w_a$ is not degenerate with the Hubble constant, as we can see in Fig. \ref{fig:tri-log}. 

When BAO data are added to Planck (Planck 2015 and Planck 2018), we find that $H_0$ is lowered together with improved error bars. In fact, the mean values of $H_0$ for this model are even lower than the estimated values of $H_0$ within the $\Lambda$CDM background by Planck. As one can notice, for this parametrization, $H_0 = 64.8_{-2.2}^{+2.1}$ km/s/Mpc (68\% CL) for Planck 2015+BAO and $H_0  =  64.8\pm2.1$ km/s/Mpc (68\% CL) for Planck 2018+R19 which significantly worsens the tension on $H_0$ at $3.7\sigma$. In this case $w_0$ is in the quintessence regime, i.e. $w>-1$, at about 95\% CL and we can bound $w_a$ at 68\% CL. The addition of the BAO data introduces a correlation between $H_0$ and $w_a$, that was not present in the Planck alone case, as we can see in Fig. \ref{fig:tri-log}.

Finally, if we consider again the addition of the local measurements of $H_0$ to Planck 2015 and 2018, we see that for both the cases the Hubble constant is now completely determined by the R18 and R19 priors respectively. Even in this scenario $w_0$ is in the quintessence regime at 68\% CL, completely in agreement with Planck + BAO, but for $w_a$ we have just an upper limit at 68\% CL. The addition of the Hubble constant priors, breaks the correlations present between $H_0$ and $w_0$, as we can see in Fig. \ref{fig:tri-log}.

\begingroup
\squeezetable
\begin{center}
\begin{table*}
\scalebox{0.9}{
\begin{tabular}{c|ccc|ccc}
\hline\hline
Parameters & Planck 2015 & Planck 2015+BAO & Planck 2015+R18 & Planck 2018 & Planck 2018+BAO & Planck 2018+R19 \\ \hline
$\Omega_ch^2$&
$0.1190_{-0.0014-0.0027}^{+0.0014+0.0028}$&
$0.1194_{-0.0014-0.0026}^{+0.0014+0.0027}$&
$0.1191_{-0.0014-0.0029}^{+0.0014+0.0027}$&
$0.1198_{-0.0013-0.0026}^{+0.0013+0.0027}$&
$0.1203_{-0.0013-0.0025}^{+0.0013+0.0026}$&
$0.1198_{-0.0013-0.0026}^{+0.0013+0.0026}$
\\
$\Omega_bh^2$&
$0.02229_{-0.00016-0.00031}^{+0.00016+0.00031}$&
$0.02224_{-0.00015-0.00029}^{+0.00015+0.00030}$&
$0.02228_{-0.00015-0.00030}^{+0.00015+0.00031}$&
$0.02241_{-0.00015-0.00029}^{+0.00015+0.00029}$&
$0.02236_{-0.00014-0.00028}^{+0.00014+0.00028}$&
$0.02240_{-0.00015-0.00029}^{+0.00015+0.00030}$
\\
$100\theta_{MC}$&
$1.04081_{-0.00032-0.00066}^{+0.00033+0.00065}$&
$1.04074_{-0.00032-0.00064}^{+0.00033+0.00065}$&
$1.04081_{-0.00033-0.00063}^{+0.00033+0.00064}$&
$1.04095_{-0.00031-0.00060}^{+0.00030+0.00060}$&
$1.04089_{-0.00030-0.00060}^{+0.00030+0.00059}$&
$1.04096_{-0.00030-0.00060}^{+0.00030+0.00058}$
\\
$\tau$&
$0.074_{-0.017-0.034}^{+0.017+0.035}$&
$0.077_{-0.017-0.034}^{+0.017+0.033}$&
$0.075_{-0.017-0.034}^{+0.017+0.033}$&
$0.0539_{-0.0075-0.015}^{+0.0076+0.016}$&
$0.0534_{-0.0076-0.016}^{+0.0076+0.015}$&
$0.0540_{-0.0075-0.015}^{+0.0076+0.016}$
\\
$n_s$&
$0.9668_{-0.0045-0.0090}^{+0.0045+0.0087}$&
$0.9655_{-0.0044-0.0087}^{+0.0044+0.0085}$&
$0.9666_{-0.0045-0.0088}^{+0.0046+0.0090}$&
$0.9658_{-0.0043-0.0085}^{+0.0042+0.0085}$&
$0.9645_{-0.0042-0.0081}^{+0.0043+0.0081}$&
$0.9659_{-0.0042-0.0082}^{+0.0042+0.0085}$
\\
${\rm{ln}}(10^{10}A_s)$&
$3.081_{-0.034-0.067}^{+0.034+0.067}$&
$3.087_{-0.033-0.064}^{+0.034+0.065}$&
$3.083_{-0.033-0.066}^{+0.034+0.064}$&
$3.043_{-0.015-0.032}^{+0.016+0.031}$&
$3.043_{-0.016-0.033}^{+0.016+0.032}$&
$3.043_{-0.015-0.031}^{+0.016+0.032}$
\\
$w_0$&
$-1.06_{-0.55-0.76}^{+0.35+0.87}$&
$-0.64_{-0.25-0.39}^{+0.18+0.43}$&
$-0.75_{-0.20-0.44}^{+0.30+0.39}$&
$-1.04_{-0.55-0.76}^{+0.36+0.88}$&
$-0.61_{-0.25-0.42}^{+0.21+0.44}$&
$-0.75_{-0.17-0.44}^{+0.30+0.37}$
\\
$w_a$&
$unconstrained$&
$-0.82_{-0.33}^{+0.60}>-1.69$&
$<-1.19,\;  \mbox{unconst.} $&
$<-1.25, \; \mbox{unconst.} $&
$-0.94_{-0.42}^{+0.60}>-1.85$&
$<-1.31,\; \mbox{unconst.}$
\\
$\Omega_{m0}$&
$0.219_{-0.082-0.097}^{+0.030+0.136}$&
$0.340_{-0.025-0.043}^{+0.020+0.044}$&
$0.263_{-0.013-0.023}^{+0.011+0.024}$&
$0.221_{-0.082-0.096}^{+0.029+0.138}$&
$0.343_{-0.025-0.045}^{+0.022+0.046}$&
$0.261_{-0.010-0.020}^{+0.010+0.022}$
\\
$\sigma_8$&
$0.96_{-0.07-0.18}^{+0.12+0.15}$&
$0.809_{-0.024-0.045}^{+0.024+0.047}$&
$0.887_{-0.020-0.040}^{+0.020+0.040}$&
$0.94_{-0.064-0.17}^{+0.12+0.15}$&
$0.795_{-0.021-0.041}^{+0.021+0.042}$&
$0.877_{-0.016-0.033}^{+0.016+0.034}$
\\
$H_0\,{\rm [km/s/Mpc]}$&
$83_{-8}^{+15} >62$&
$64.8_{-2.2-4.0}^{+2.1+4.1}$&
$73.6_{-1.6-3.2}^{+1.6+3.1}$&
$83_{-8}^{+15} >62$&
$64.8_{-2.1-4.1}^{+2.1+4.2}$&
$74.1_{-1.4-2.8}^{+1.4+2.9}$
\\
\hline\hline
\end{tabular}
}
\caption{Observational constraints on various free and derived parameters at 68\% and 95\% CL in the cosmological model where DE follows the Logarithmic parametrization, for various observational data and their combinations.  }
\label{tab:Log}
\end{table*}
\end{center}
\endgroup

\begingroup
\squeezetable
\begin{center}
\begin{table*}
\scalebox{0.9}{
\begin{tabular}{c|ccc|ccc}
\hline\hline
Parameters & Planck 2015 & Planck 2015+BAO & Planck 2015+R18 & Planck 2018 & Planck 2018+BAO & Planck 2018+R19 \\ \hline
$\Omega_ch^2$&
$0.1190_{-0.0014-0.0027}^{+0.0014+0.0027}$&
$0.1194_{-0.0013-0.0026}^{+0.0013+0.0027}$&
$0.1190_{-0.0014-0.0027}^{+0.0014+0.0027}$&
$0.1197_{-0.0013-0.0027}^{+0.0014+0.0027}$&
$0.1203_{-0.0013-0.0025}^{+0.0013+0.0025}$&
$0.1198_{-0.0013-0.0026}^{+0.0014+0.0028}$
\\
$\Omega_bh^2$&
$0.02229_{-0.00016-0.00030}^{+0.00016+0.00031}$&
$0.02225_{-0.00015-0.00030}^{+0.00015+0.00030}$&
$0.02229_{-0.00015-0.00029}^{+0.00015+0.00030}$&
$0.02241_{-0.00015-0.00030}^{+0.00015+0.00030}$&
$0.02235_{-0.00014-0.00028}^{+0.00015+0.00028}$&
$0.02241_{-0.00015-0.00029}^{+0.00015+0.00028}$
\\
$100\theta_{MC}$&
$1.04080_{-0.00031-0.00064}^{+0.00031+0.00063}$&
$1.04075_{-0.00032-0.00063}^{+0.00032+0.00062}$&
$1.04081_{-0.00032-0.00064}^{+0.00031+0.00063}$&
$1.04096_{-0.00031-0.00062}^{+0.00031+0.00061}$&
$1.04089_{-0.00031-0.00059}^{+0.00031+0.00061}$&
$1.04094_{-0.00031-0.00061}^{+0.00031+0.00060}$
\\
$\tau$&
$0.074_{-0.018-0.034}^{+0.018+0.034}$&
$0.078_{-0.017-0.034}^{+0.017+0.034}$&
$0.075_{-0.018-0.032}^{+0.017+0.034}$&
$0.0539_{-0.0075-0.015}^{+0.0076+0.016}$&
$0.0533_{-0.0074-0.016}^{+0.0075+0.016}$&
$0.0536_{-0.0075-0.015}^{+0.0075+0.016}$
\\
$n_s$&
$0.9666_{-0.0046-0.0090}^{+0.0046+0.0088}$&
$0.9659_{-0.0044-0.0089}^{+0.0044+0.0086}$&
$0.9666_{-0.0045-0.0088}^{+0.0044+0.0089}$&
$0.9659_{-0.0044-0.0087}^{+0.0044+0.0087}$&
$0.9644_{-0.0041-0.0078}^{+0.0041+0.0083}$&
$0.9655_{-0.0043-0.0087}^{+0.0043+0.0083}$
\\
${\rm{ln}}(10^{10}A_s)$&
$3.081_{-0.035-0.067}^{+0.035+0.067}$&
$3.089_{-0.033-0.066}^{+0.033+0.064}$&
$3.081_{-0.033-0.065}^{+0.033+0.066}$&
$3.043_{-0.016-0.031}^{+0.016+0.032}$&
$3.043_{-0.015-0.032}^{+0.015+0.032}$&
$3.043_{-0.015-0.031}^{+0.016+0.031}$
\\
$w_0$&
$-1.02_{-0.56}^{+0.46}, {\mbox unconst.}$&
$-0.68_{-0.24-0.48}^{+0.24+0.47}$&
$-0.64_{-0.23-0.70}^{+0.46+0.57}$&
$-0.99_{-0.54}^{+0.50}, {\mbox unconst.}$&
$-0.66_{-0.24-0.49}^{+0.25+0.46}$&
$-0.65_{-0.21-0.67}^{+0.44+0.54}$
\\
$w_a$&
$<-0.88 <0.27$&
$-0.53_{-0.35-0.76}^{+0.40+0.71}$&
$<-1.09 <0.21$&
$<-1.08 <0.18$&
$-0.61_{-0.38-0.74}^{+0.37+0.73}$&
$<-1.19 <0.08$
\\
$\Omega_{m0}$&
$0.214_{-0.075-0.09}^{+0.027+0.12}$&
$0.337_{-0.026-0.049}^{+0.025+0.051}$&
$0.263_{-0.013-0.023}^{+0.012+0.025}$&
$0.215_{-0.074-0.09}^{+0.026+0.12}$&
$0.338_{-0.026-0.049}^{+0.026+0.051}$&
$0.261_{-0.010-0.019}^{+0.010+0.021}$
\\
$\sigma_8$&
$0.97_{-0.06-0.16}^{+0.11+0.14}$&
$0.811_{-0.026-0.048}^{+0.025+0.053}$&
$0.887_{-0.020-0.041}^{+0.021+0.040}$&
$0.95_{-0.06-0.16}^{+0.11+0.14}$&
$0.798_{-0.026-0.044}^{+0.022+0.050}$&
$0.878_{-0.017-0.031}^{+0.017+0.032}$
\\
$H_0\,{\rm [km/s/Mpc]}$&
$84_{-8}^{+14} >65$&
$65.1_{-2.6-4.8}^{+2.3+4.9}$&
$73.6_{-1.7-3.3}^{+1.7+3.3}$&
$83_{-8}^{+15} >64$&
$65.2_{-2.8-4.8}^{+2.2+5.1}$&
$74.1_{-1.4-2.7}^{+1.4+2.8}$
\\
\hline\hline
\end{tabular}
}
\caption{Observational constraints on various free and derived parameters at 68\% and 95\% CL in the cosmological model where DE obeys the BA parametrization, for various observational data and their combinations. }
\label{tab:BA}
\end{table*}
\end{center}
\endgroup

\begin{figure*}
    \centering
    \includegraphics[width=0.65\textwidth]{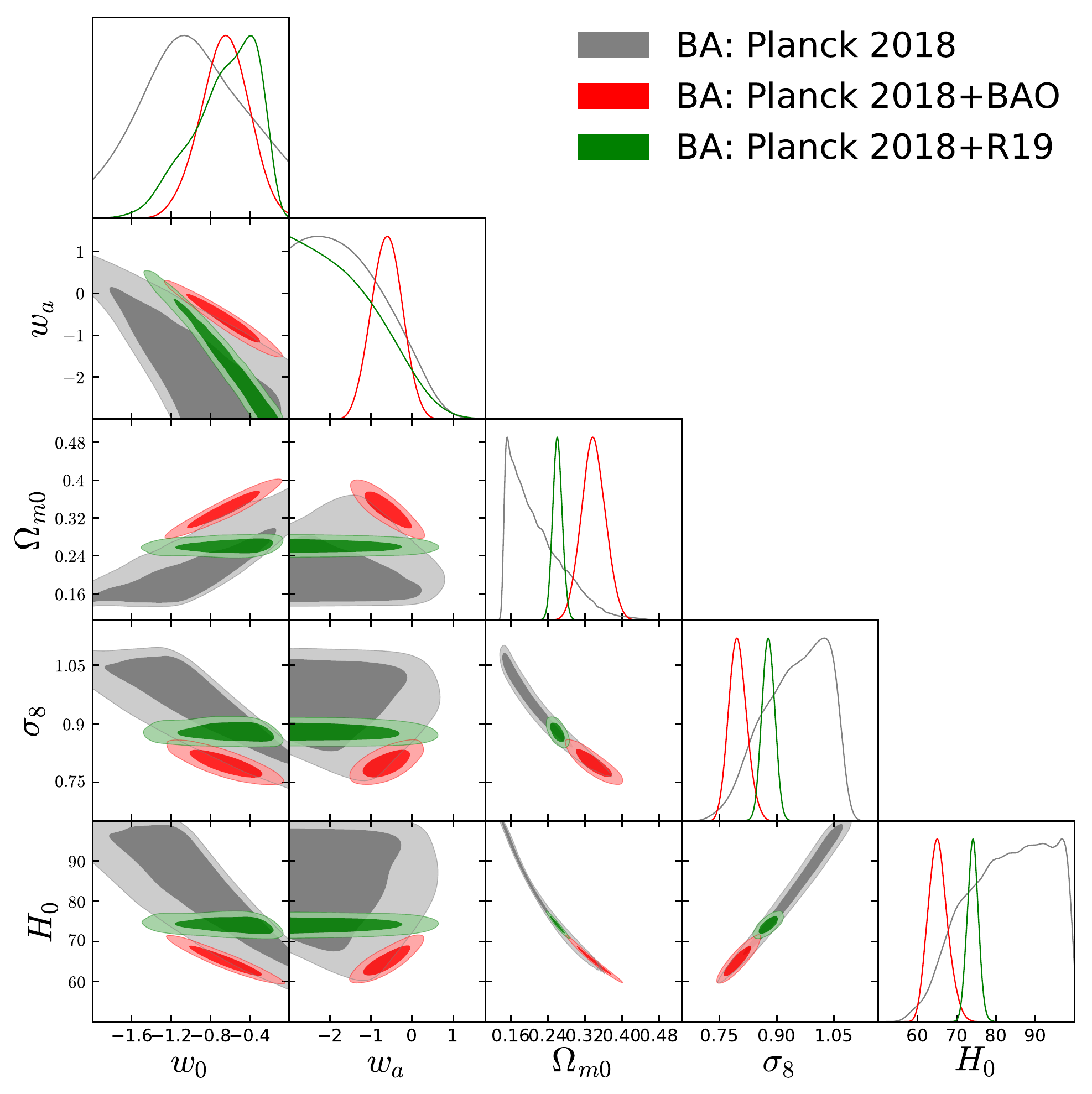}
    \caption{We show the 1-dimensional marginalized posterior distributions and 2-dimensional contour plots of some key parameters of the BA cosmological model using Planck 2018 and its combination with BAO and R19.  }
     \label{fig:tri-BA}
\end{figure*}

\subsection{BA parametrization}
\label{sec-ba}

Finally, we come to the last model in this article, namely the Barboza-Alcaniz parametrization of eqn. (\ref{ba}). We have similarly constrained this model using the same observational data described above and the results are shown in Table~\ref{tab:BA}. We have on the first 3 columns the constraints on the model from Planck 2015 data, Planck 2015+BAO and Planck 2015+R18, while on the 4th to 6th columns the constraints on the model from Planck 2018 data, Planck 2018+BAO and Planck 2018+R19. In fact, also for the BA model Planck is no more in tension with R18 and R19. Moreover, in Fig.~\ref{fig:tri-BA} we have the triangular plot for Planck 2018 and its combination with BAO and R19.

Our analyses show that Planck alone cannot give tight constraints on the parameters, as in the previous models, giving for $H_0$ just a lower limit at 95\% CL. However, like in the Log case, $w_0$ is completely in agreement within $1\sigma$ with a cosmological constant $w=-1$ but we have an upper limit for $w_a$ also at 95\% CL, like in the CPL case. 

The addition of BAO restores the Hubble tension. In fact, the inclusion of BAO to Planck 2015 and Planck 2018 lowers the values of $H_0$, that is smaller than the $\Lambda$CDM based Planck's estimation, but with slightly larger error bars. In fact, for this BA parametrization, $H_0 = 65.1_{-2.6}^{+2.3}$ km/s/Mpc (68\% CL) for Planck 2015+BAO and $H_0  =  65.2_{-2.8}^{+2.2}$ km/s/Mpc (68\% CL) for Planck 2018+R19, restoring the tension on $H_0$ at $3.4$ standard deviations. In this case $w_0$ is again in the quintessence regime $w>-1$ at more than 68\% CL, and we can bound $w_a$, that will be negative at more 68\% CL for both Planck 2015+R18 and Planck 2018+R19. In particular we find $w_a = -0.53_{-0.35}^{+0.40}$ at 68\% CL for Planck 2015+BAO and $w_a = -0.61_{-0.38}^{+0.37}$ at 68\% CL for Planck 2018+R19. With the addition of the BAO data $H_0$ shows a correlation with both the dark energy equation of state parameters ($w_0$, $w_a$), that was not present in the Planck alone case, as we can see in Fig. \ref{fig:tri-BA}.

Finally, as already observed with other DE parametrizations, the inclusion of the local measurements of $H_0$ to Planck 2015 and 2018, fixes the Hubble constant values to the R18 and R19 priors respectively. Even in this scenario $w_0$ is in the quintessence regime, completely in agreement with Planck + BAO, but with larger error bars. Regarding instead $w_a$, we have just upper limits at 68\% CL and 95\% CL, with the negative value preferred at about $2\sigma$ for Planck 2018+R19.

\begin{figure*}
    \centering
    \includegraphics[width=0.85\textwidth]{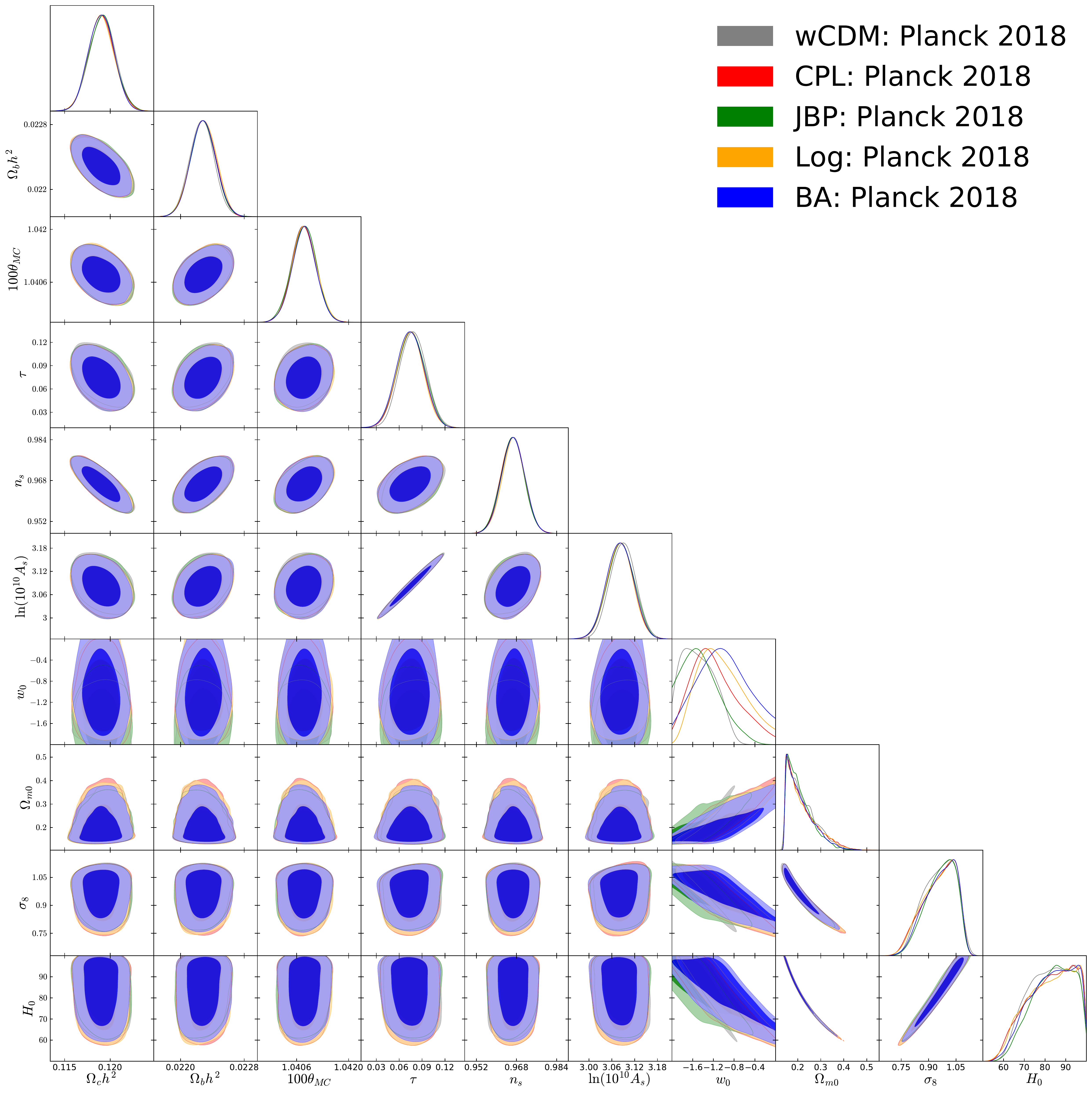}
    \caption{We show the 1-dimensional marginalized posterior distributions and 2-dimensional contour plots of all the cosmological parameters for all the dark energy models considered here for the Planck 2018 dataset.  }
     \label{fig:comparison}
\end{figure*}
\begingroup                      
\begin{center}                                                                                                                  
\begin{table*}                                                                                                                
\begin{tabular}{ccc}                                                                                                            
\hline\hline                                                                                                                    
$\ln B_{ij}$ & ~~~~~~~Strength of evidence for model ${M}_i$ \\ \hline
$0 \leq \ln B_{ij} < 1$ & Weak \\
$1 \leq \ln B_{ij} < 3$ & Definite/Positive \\
$3 \leq \ln B_{ij} < 5$ & Strong \\
$\ln B_{ij} \geq 5$ & Very strong \\
\hline\hline                                                                                                                    
\end{tabular}                                                                                                                   
\caption{Summary of the revised Jeffreys scale by Kass and Raftery as in Ref.~\cite{Kass:1995loi}. } \label{tab:jeffreys}                                            
\end{table*}                                                
\end{center}                                                                                                                    
\endgroup 
\begingroup                                                                                                                     
\begin{center}                                              
\begin{table*}                                              
\begin{tabular}{|cccc|c|}                                    
\hline                                                
Dataset & Model & $\ln B_{ij}$ & ~~~~~~~Strength of evidence \\ \hline

CMB & $w$CDM & 0.7 & Weak for $w$CDM\\
CMB+BAO & $w$CDM & -0.1 & Weak for $\Lambda$CDM \\
CMB+R19 & $w$CDM & 4.6 & Strong for $w$CDM\\
\hline
CMB & CPL & 3.5 & Strong for CPL\\
CMB+BAO & CPL & 1.6 & Definite for CPL\\
CMB+R19 & CPL & 5.5 & Very Strong for CPL\\
\hline
CMB & JBP & 2.7 & Definite for JBP\\
CMB+BAO & JBP & 1.1 & Definite for JBP\\
CMB+R19 & JBP & 6.2 & Very Strong for JBP\\
\hline
CMB & Log & 2.2 & Definite for Log\\
CMB+BAO & Log & 1.7 & Definite for Log\\
CMB+R19 & Log & 5.6 & Very Strong for Log\\
\hline
CMB & BA & 2.3 & Definite for BA\\
CMB+BAO & BA & 0.9 & Weak for BA\\
CMB+R19 & BA & 6.6 & Very Strong for BA\\

\hline                                                
\end{tabular}
\caption{Values of $\ln B_{ij}$ and the strength of the evidence for the Dynamical Dark Energy models considered here against the $\Lambda$CDM scenario, as obtained in our analysis for different dataset combinations. The negative sign indicates that the $\Lambda$CDM is preferred over the DDE model and vice versa. }\label{tab:bayesian}                                                                                            \end{table*}                                                    \end{center}                                                
\endgroup

\subsection{Model Comparison}
\label{BE}

Let us now perform a comparison between the simplest constant dark energy equation of state, the $w$CDM model, and several famous parametrizations of the dark energy equation of state parameters varying with the redshift. The first thing we can see is that the constraints on the cosmological parameters, both free and derived, are almost unaltered by the choice of the DE parametrization. One can effectively visualize this in Fig.~\ref{fig:comparison}, where we compare the 1D marginalized posterior distributions and 2D contour plots of all the cosmological parameters for all the dark energy parametrizations considered here (for the Planck 2018 dataset alone). We can conclude that the only difference we have between the models, is the different direction of correlation in the plane $w_0-w_a$, that is directly related to the difference in the parametrization, but all of them agree in ruling out the cosmological constant at 2 standard deviations.

As it is evident from the Tables in the previous sections (i.e. Tables~\ref{tab:wCDM}, \ref{tab:CPL}, \ref{tab:JBP}, \ref{tab:Log}, \ref{tab:BA}) and in particular from Fig.~\ref{fig2}, considering the Planck 2018 data only, a model having the dark energy equation of state free to vary can alleviate in an excellent way the Hubble constant tension, at the price of behaving as phantom dark energy. However, according to the results obtained here, the solution of the $H_0$ tension is not supported by the BAO data. In fact the addition of these measurements restore the $H_0$ tension at about $2.7$ standard deviations for the CPL and JBP parametrization, and above 3$\sigma$ for the $w$CDM, Log and BA cases.

However, the problem with BAO data regarding the solution to the $H_0$ tension is model dependent. That means, BAO data may work better with other DE models. For example, a class of dynamical DE models having only one free parameter  can significantly alleviate the $H_0$ tension even if BAO are added to CMB (i.e. for the combination CMB+BAO) {\color{blue}\cite{Yang:2018qmz}}. This is not the only report in the literature. The same conclusion holds in the recently introduced Phenomenologically Emergent Dark Energy (PEDE) model, a dynamical DE model {\color{blue}\cite{Li:2019yem,Pan:2019hac,Yang:2020myd}}. In particular, within this  PEDE scenario, $H_0 = 71.55^{+0.55}_{-0.57}$ (68\% CL for Planck 2015 + BAO) {\color{blue}\cite{Pan:2019hac}} and $H_0= 71.54^{+0.53}_{-0.55}$ (68\% CL for Planck 2018 + BAO) {\color{blue}\cite{Yang:2020myd}}. 
Thus, the solution to the $H_0$ tension can be supported well by the BAO data in other dynamical DE models. 
At this point it might be worth remembering that interacting DE models also act as an excellent choice for alleviating the $H_0$ tension. Although one could recast an interacting DE model as a non-interacting scenario where both dark matter and dark energy enjoy variable equations of state. In particular, 
interacting DE models where DE has either a constant equation-of-state or a variable equation-of-state, have been found to alleviate the $H_0$ tension quite significantly, see for instance {\color{blue}\cite{Kumar:2016zpg,Kumar:2017dnp,DiValentino:2017iww,Yang:2018euj,Yang:2018uae,Kumar:2019wfs,Yang:2019uzo,Pan:2019gop,Yang:2019uog,Pan:2020bur, Pan:2019jqh,Martinelli:2019dau,DiValentino:2019ffd,DiValentino:2019jae,Pan:2020bur}}. In Ref. {\color{blue}\cite{DiValentino:2017iww}}, the authors showed that within a simple interaction function $Q \propto \rho_x$ characterizing the energy transfer between DE (irrespective of its equation of state, $w_x$) and DM, the Hubble constant assumes higher values compared to its low estimation within the $\Lambda$CDM cosmology by Planck team. According to Ref. {\color{blue}\cite{DiValentino:2017iww}}, interacting DE with $w_x = -1$,  the estimated values of the Hubble constant are: $H_0 = 68.9 \pm 1.2$ (68\% CL, Planck 2015 alone) and $H_0 = 68.58^{+0.8}_{-1.1}$ (68\% CL, Planck 2015 + BAO). For a constant equation of state in DE, $w_x$ ($w_x \neq -1$) within the same interaction framework, $H_0$ assumes higher values for both Planck 2015 and Planck 2015 + BAO {\color{blue}\cite{DiValentino:2017iww}}.  The conclusion remains same (for the same interaction function $Q \propto \rho_x$) for the Planck 2018 and Planck 2018 + BAO as well as reported in  {\color{blue}\cite{DiValentino:2019ffd,DiValentino:2019jae}} showing that the inclusion of BAO data works well to alleviate the $H_0$ problem within $3\sigma$. Interestingly, for the dynamical dark energy equation of state parameters, the $H_0$ tension is significantly alleviated {\color{blue}\cite{Yang:2018uae,Pan:2019gop}}. 
Moreover, for other interaction functions, the tension on $H_0$ is equally alleviated {\color{blue}\cite{Yang:2018euj,Pan:2019jqh,Pan:2020bur}}.  In particular, in {\color{blue}\cite{Pan:2019jqh}}, the interaction model allows a sign changeable feature characterized by the direction of energy flow between the dark sectors. 
Unlike the dynamical DE models of this work where the $H_0$ solution is not supported by the  BAO data, in the context of other parametrized DE models and also in the framework of interacting scenarios,  this does not appear to be a serious problem. However, the selection of the models remains an important issue although.

Finally, out of the present five parametrized DE models, in order to appreciate which dynamical DE model of this series performs better the fitting of the CMB data and the resolution of the $H_0$ tension at the same time, we follow a Bayesian evidence analysis and compute the values through which we estimate the fitting of the cosmological model. 

For the Bayes theorem the posterior probability of the parameter $\theta$, given a particular data set $x$ and considering a model $M$, is obtained as:
\begin{eqnarray}\label{BE}
p(\theta|x, M) = \frac{p(x|\theta, M)\,\pi(\theta|M)}{p(x|M)},
\end{eqnarray}
where $p(x|\theta, M)$ refers to the likelihood function, $\pi(\theta|M)$ denotes the prior information, and $p(x|M)$ the Bayesian evidence defined as follows

\begin{eqnarray}\label{sp-be01}
E \equiv p(x|M) = \int d\theta\, p(x|\theta,M)\,\pi(\theta|M). 
\end{eqnarray}

We can compute the ratio of Evidences, called the Bayes factor of the considered model $M_i$ compared to the reference model $M_j$, like:

\begin{eqnarray}
B_{ij} = \frac{p(x| M_i)}{p(x|M_j)}
\end{eqnarray}
and it indicates how the observational data support the model $M_i$, that is under consideration, over the reference model $M_j$ (here the $\Lambda$CDM model). If the Bayes factor is positive, then the data prefer the model $M_i$ to the model $M_j$. The evidence for model $M_i$ against the model $M_j$ is then quantified using the revised Jeffreys scale by Kass and Raftery~{\color{blue}\cite{Kass:1995loi}} showed 
in Table~\ref{tab:jeffreys}. 

In order to compute the Bayesian evidence, we use the publicly available cosmological package \texttt{MCEvidence}\footnote{\href{https://github.com/yabebalFantaye/MCEvidence}{github.com/yabebalFantaye/MCEvidence}. See also {\color{blue}\cite{Heavens:2017hkr,Heavens:2017afc}}.}. 
In Table \ref{tab:bayesian} we present the $\ln B_{ij}$ values of the DE models considered here with respect to the base $\Lambda$CDM scenario. The negative values of $\ln B_{ij}$ are indicating the preference of the data for the $\Lambda$CDM scenario, while the positive values denote the opposite scenario. Looking at the numerical values of $\ln B_{ij}$ and considering the Jeffreys scale as shown in Table \ref{tab:jeffreys}, we may clearly conclude that the DE models are always preferred over the $\Lambda$CDM model for all the datasets combination considered in this work.

Given the fact that all the DE models in this work are resolving the $H_0$ tension exceptionally well, to understand which model performs better the fitting of the CMB data we can compare the $\ln B_{ij}$. We find that in general, despite having one more degree of freedom, dynamical DE models improve significantly the fit of the CMB data and are preferred with respect to the $w$CDM model. Moreover, among all the redshift dependent dark energy equation of state parameters, the best in fitting the CMB data and solving the Hubble constant tension at the same time is the CPL parametrization. Moreover, in this case, the addition of the BAO data keeps the Hubble constant tension below $3\sigma$, therefore in agreement with a possible statistical fluctuation due to the method used to extract the BAO measurements (i.e. assuming a cosmological constant and no DE perturbations).

\section{Summary and concluding remarks}
\label{sec-discuss}

The physics of dark energy is an obscure chapter for modern cosmology. If we try to be precise, then almost 20 years have passed since the detection of some dark energy-like component accelerating the present expansion of our universe, and still we do not know the nature of dark energy. The simplest explanation has the name of $\Lambda$CDM cosmology, where $\Lambda>0$ is inserted forcefully within the framework of Einstein's general theory of relativity, and has been found to be a successful model to trace this late cosmic acceleration. However, investigations since the introduction of $\Lambda$,  have raised many controversies. The cosmological constant problem, related to its estimation from both theoretical and observational grounds, is one of such biggest mysteries leading to many works. On the other hand, the cosmic coincidence problem, related to the order of dark matter and dark energy densities at the current epoch of the cosmic evolution has been another inspiration to look for alternative cosmological models beyond $\Lambda$CDM.  The search for alternative cosmological models was triggered when the estimations of $H_0$ within $\Lambda$CDM model were found to mismatch between different observational missions at more than $4$ standard deviations, see for instance~{\color{blue}\cite{Riess:2019cxk,Wong:2019kwg,Camarena:2019moy}}. 
A possible simple approach to alleviate the above problems might be the assumption of a dark energy state parameter other than $w = -1$, referring to the cosmological constant fluid. 

Thus, in the present work we have considered a number of dark energy models in which the dark energy equation of state parameter can be different from $-1$. Precisely, we have considered two possibilities, one where the dark energy equation of state is constant and on the other hand where it is dynamically evolving with the redshift. In the context of dynamical dark energy equation of state, we have considered 
a variety of well known models. The motivation to consider these well known models is to investigate the $H_0$ tension in light of the latest CMB measurements released from Planck 2018. Along with the recent CMB measurements from Planck 2018, we have considered BAO and local measurements of $H_0$, as measured by R18 and R19. Moreover, we have included the CMB data from Planck 2015 measurements in order to check the improvements of the cosmological constraints during the years.

In Tables \ref{tab:wCDM}, \ref{tab:CPL}, \ref{tab:JBP}, \ref{tab:Log}, and \ref{tab:BA} we have summarized our results. Moreover, 
we displayed the triangular plots in Figs. \ref{fig:tri-wcdm}, \ref{fig:tri-cpl}, \ref{fig:tri-jbp}, \ref{fig:tri-log} and \ref{fig:tri-BA}, respectively for $w$CDM, CPL, JBP, Log and BA models for Planck 2018, Planck 2018+BAO and Planck 2018+R19.

Our analyses show that Planck alone returns very high mean values of the Hubble constant, almost unconstrained at 95\% CL,  and consequently, the Hubble constant tension for all the DE models considered in this work is solved. We refer to Fig.~\ref{fig2} to have a look at the estimations of $H_0$ for all the models and datasets considered here.  
In summary, one can claim that the dynamical DE models might be considered to be an effective way to solve/alleviate the $H_0$ tension. However, the 
inclusion of BAO data to Planck restores the $H_0$ tension at about $2.7$ standard deviations for the CPL and JBP parametrizations, and above 3$\sigma$ for the $w$CDM, Log and BA cases. Therefore, these dynamical DE models are disfavoured in a more general scenario. In fact, in the literature there are other possibilities that can solve the $H_0$ tension even when adding BAO data (see for example {\color{blue}\cite{Pan:2019hac,DiValentino:2020naf}}).

\section*{Acknowledgments}
The authors express their sincere thanks to the referee for her/his essential comments that improved the quality of the manuscript.  WY work is supported by the National Natural Science Foundation of China under Grants No. 11705079 and No. 11647153. EDV acknowledges support from the European Research Council in the form of a Consolidator Grant with number 681431.
SP was supported by the Mathematical Research Impact-Centric Support Scheme (MATRICS), File No. MTR/2018/000940, given by the Science and Engineering Research Board (SERB), Govt. of India. 
YW is supported by the National Natural Science Foundation of China under Grant No. 11575075.
JL is supported by the National Natural Science Foundation of China under Grant No. 11645003.

\section*{Data Availability}
We have used the standard cosmological probes such as cosmic microwave background radiation  from Planck 2015 and Planck 2018, baryon acoustic oscillations distance
measurements from various astronomical missions, and measurements of the Hubble constant from Hubble Space Telescope. All of them are freely available.

\bibliography{references}
\end{document}